\documentclass{aa}
\usepackage{epsf,graphicx}

\begin{document}


\title{ Thermal Conductivity of Neutrons
            in Neutron Star Cores}

\author{D.A.\ Baiko\inst{1}, P.\ Haensel\inst{2}
        \and D.G.\ Yakovlev\inst{1,3}}
\offprints{P. Haensel}

\institute{
        A.F.\ Ioffe Physical Technical Institute,
        Politekhnicheskaya 26, 194021, St.Petersburg, Russia 
        \and
        N.\ Copernicus Astronomical Center,
        Bartycka 18,  00-716, Warsaw, Poland
        \and
        Institute for Theoretical Physics,
        University of California, Santa Barbara,
        CA 93106, USA\\
{\email 
 baiko@mail.rit.edu, haensel@camk.edu.pl, yak@astro.ioffe.rssi.ru}
         }
\date{Received; accepted}
\authorrunning{D.A. Baiko et al.}
\titlerunning{Thermal conductivity of neutrons in neutron
star cores}

\newcommand{\dd}{\mbox{d}}                     


\abstract{
The diffusive thermal conductivity of neutrons in dense
matter [$\rho \sim (1 \,- \, 8) \times 10^{14}$ g cm$^{-3}$]
of neutron star cores is calculated. The contribution from
neutron--neutron and neutron--proton collisions is
taken into account. 
We use the transition probabilities 
calculated for symmetric dense nucleon matter on the basis
of the Dirac--Brueckner approach to the in-medium effects
and the Bonn model of bare nucleon--nucleon interaction.  
The diffusive thermal conductivity of neutrons
in the presence of neutron and proton superfluidities
is analyzed in a microscopic manner; 
the effects of superfluidity are shown to
be significant. The low temperature behavior of the
thermal conductivity appears to be extremely sensitive
to the relation between critical temperatures of neutrons and protons.
The results
are fitted by simple analytic expressions.
In combination with the formulae
for the electron and muon thermal conductivities, obtained earlier,
the present expressions provide 
a realistic description of the full
diffusive thermal conductivity in the neutron star cores
for normal and various superfluid phases.
}
\maketitle

\section{Introduction}
\label{s1}
We study the thermal conductivity
of neutrons in the neutron star cores,
in the density range $\rho \sim (0.5 \, - \, 3) \rho_0$, where
$\rho_0 = 2.8 \times 10^{14}$ g cm$^{-3}$
is the normal nuclear density.
The problem is important 
for numerical simulations
of cooling of young neutron stars, before internal thermal
equilibrium is achieved.
The non-relaxed stage lasts $10-100$ yr
(Van Riper, \cite{VR91}, Umeda et al., \cite{Uetal93}, Lattimer et al.,
\cite{lvpp94}) depending on the model of dense matter.

At densities close to $\rho_0$, neutron star matter is
expected to consist mostly of neutrons ($n$), with a small
(a few percent) admixture of protons ($p$), electrons ($e$)
and, at densities where the electron Fermi energy exceeds the
muon rest energy, also of muons ($\mu$). In
what follows, we adopt this $npe\mu$ model
of neutron star matter within the considered density range.
At higher $\rho$ other particles may appear, first of all hyperons
(e.g., Balberg et al.\ \cite{blc99}).
We discuss briefly thermal conductivity of hyperonic matter in Sect.\ 4.

Numerical simulations of neutron star
cooling traditionally
employ the thermal conductivity obtained by Flowers \&
Itoh (\cite{FI79,FI81}) about 20 years  ago. These  authors
proposed
an analytic fit valid for non-superfluid matter,
and discussed qualitatively
the effects of nucleon superfluidity. It is now
generally recognized that the neutron star cores
can be in a superfluid state (see, e.g.,
Pines, \cite{P91}, Page \& Applegate, \cite{PA92}, 
Yakovlev et al., \cite{yls99}).
Moreover, Flowers \& Itoh (\cite{FI79,FI81}) calculated the thermal
conductivity for one specific model of dense matter
which is inconvenient as the equation of state
of neutron star cores is generally unknown.
The thermal conductivity of the neutron star cores
was analyzed also by Wambach et al.\
(\cite{Wetal93}) and
by Sedrakian et al.\ (\cite{Setal94}),
again for some selected models of dense matter
and neglecting the superfluidity effects.
Thus, it is desirable to obtain new
expressions which apply to various models of matter and
take into account nucleon superfluidity.
In contrast to the  studies
of Flowers \& Itoh (\cite{FI79}), Wambach et al.\ (\cite{Wetal93}),
Sedrakian et al.\ (\cite{Setal94}), 
we consider the effects of nucleon
superfluidity 
in a microscopic manner by evaluating the diffusive
thermal conductivity of neutrons in superfluid matter.
  
The main heat carriers in the neutron star cores are
neutrons (the most abundant particles) as well as
electrons and muons 
which have large mean free paths since they 
experience only relatively weak, Coulomb 
interactions. 
The thermal conductivity of protons
is small (Flowers \& Itoh, \cite{FI79}) and can be neglected.
The heat conduction of electrons and muons is
limited by Coulomb collisions
with $e$, $p$ and $\mu$,
while that of  neutrons is limited by their collisions
with $n$ and $p$
due to the strong nucleon--nucleon interaction. Accordingly,
the neutron transport is almost independent of
the electron and muon one,
and can be studied separately. This property
was proved by Flowers \& Itoh (\cite{FI79})
who decoupled the overall heat--conduction equations into two blocks
describing the conductivities of neutrons and electrons.

The thermal conductivity of electrons and muons
was reconsidered by Gnedin \& Yakovlev (\cite{GY95})
who obtained simple expressions  
valid for a wide class of models of superfluid and non-superfluid matter.
In the present article, we reanalyze the thermal conduction
of neutrons utilizing some new developments in the 
nucleon--nucleon interaction theory.
In combination with the results of
Gnedin \& Yakovlev (\cite{GY95}), this
describes the diffusive
heat transport in not too dense neutron star cores.

In the presence of neutron superfluidity 
there may be another channel of heat transport, the so-called
convective counterflow of normal component of matter with
respect to superfluid one.
This mechanism is known to be extremely effective in superfluid helium
(e.g., Tilley \& Tilley, \cite{TT90})
but in the case of neutron star matter the situation is more complicated.
The related effects require separate study 
and will not be considered here.

\section{General Formalism}
\label{s2}
Consider a neutron star core at densities
$\rho$ from about $0.5 \rho_0$ to 3$\rho_0$.
The lower limit corresponds to the core--crust transition
(Lorenz et al., \cite{Letal93}, Pethick \& Ravenhall, \cite{PR95}), 
while the upper limit is the typical central density
of a not too massive star.
We treat neutrons as non-relativistic particles
and do not study higher $\rho$ where they
become mildly relativistic.

To calculate the neutron conductivity, it is
sufficient to consider
the nucleon component of matter. In the
density range of study,
typical Fermi energy of neutrons changes from a few ten
to a few hundred MeV (e.g., Shapiro \& Teukolsky, \cite{ST83}).
Fermi energy of protons varies
from several MeV  to a few tens of MeV. Since
the temperature in the cores of cooling neutron stars is
typically well below $10^{10}$~K (1 MeV),
both neutrons and protons form
strongly degenerate Fermi liquids.

In the limit of strong degeneracy
transport properties
of a nucleon liquid can be described using the concept
of {\it quasiparticles}. In the present section, we
restrict ourselves to the case of {\it normal} nucleon liquids
(the effects of nucleon superfluidity will be considered in Sect.\ \ref{s3}). 
Landau theory of normal Fermi liquids enables one to
identify the low temperature properties of a strongly interacting
real system with those of a dilute gas of weakly interacting
elementary excitations --- the nucleon quasiparticles (a detailed
description of the theory and its
applications can be found in Baym \& Pethick, \cite{BP91}). 
This formalism can be used effectively only for describing
small macroscopic
deviations of the real system from its
ground state.
A crucial feature of the Landau Fermi--liquid theory is the inclusion of
the quasiparticle interaction.  The kinetics of quasiparticles
can be described including only binary
quasiparticle collisions. We are interested in the collisions
of neutron quasiparticles with neutron and proton
quasiparticles. In the calculation of the neutron thermal conductivity
protons will be considered as scatterers.

\subsection{Kinetic equation}
\label{s21}
The distribution function $F_n$ of neutron quasiparticles
satisfies the Landau
kinetic equation (e.g., Baym \& Pethick, \cite{BP91}) 
which describes motion
of a given quasiparticle in a self--consistent field of its
neighbors.
However, a proper linearization reduces the Landau
equation to a Boltzmann-like equation.
The linearized expression for the quasiparticle thermal flux
density also takes its standard form.
This enables us to use
the expressions familiar from the kinetics of dilute gases
(e.g., Ziman, \cite{Z60}).
The thermal flux density of neutrons is
\begin{eqnarray}
         {\vec q}_n  & = & {2 \over (2 \pi)^3} \, \int \,
         \dd {\vec k}_n \, (\epsilon_n-\mu_n) \,
         {\vec v}_n \, F_n 
\equiv \; -
         \kappa_n \, \nabla T,
\label{q}
\end{eqnarray}
where ${\vec k}_n$, ${\vec v}_n$, $\epsilon_n$,
$\mu_n$ are the neutron wave-vector, velocity,
energy, and chemical potential, respectively;
$\kappa_n$ is the neutron thermal conductivity to be
determined, and $T$ is the temperature.
In the presence of a weak temperature
gradient, the Boltzmann equation reads
\begin{equation}
        {\vec v}_n \cdot \nabla F_n = I_{nn} +
        I_{np},
\label{kin}
\end{equation}
where $I_{nn}$ and $I_{np}$ are
the $nn$ and $np$ collision integrals:
\begin{eqnarray}
        I_{12} &=& {V^3 \over (2 \pi)^9 \, (1+\delta_{12})} \,
         \int \! \int \! \int
        \dd {\vec k}'_1 \, \dd {\vec k}_2 \,
        \dd {\vec k}'_2 
\nonumber \\
        && \times \;\sum_{\sigma'_1 \sigma_2 \sigma'_2} \, {\cal P}_{12}
         \,  \left[ F'_1 \, F'_2 \,
        (1-F_1) \, (1-F_2)  \right.
\nonumber \\ 
      && \left. - \; F_1 \, F_2 \, (1-F'_1) \,
        (1-F'_2) \right].
\label{Igen}
\end{eqnarray}
The non-primed quantities refer to the particles
before a collision event, while the primed ones refer to the
particles after the collision, $\sigma$ denotes the spin state, and
${\cal P}_{12}$ is the differential transition rate.
The rates of direct and inverse processes
are equal due to detailed balancing.
The term $\delta_{12}$ serves to avoid double counting of the final
states in the case of $nn$ collisions;
$V$ is the normalization volume.

Since the protons are treated as scatterers (but not
the heat carriers) their distribution
function coincides with the equilibrium Fermi--Dirac distribution
$f(\epsilon)$:
\begin{eqnarray}
       F_p & = &f_p \equiv f(\epsilon_p), 
\nonumber \\
         f(\epsilon)& \equiv&
       \left[1+\exp\biggl({\epsilon-\mu \over T}\biggl)
       \right]^{-1}.
\label{Fp}
\end{eqnarray}
The neutron distribution can be written as
\begin{equation}
       F_n = f_n -
       \Phi_n \, {\partial f_n \over \partial \epsilon_n},
\label{Fn}
\end{equation}
where $f_n = f(\epsilon_n)$, and
$\Phi_n = \Phi_n(\epsilon_n)$ describes
the small deviation of the distribution function from the equilibrium one.
This correction is known (Ziman, \cite{Z60}) to have the form
\begin{equation}
      \Phi_n  = -\tau_n \,
      (\epsilon_n - \mu_n) \, {{\vec v}_n
      \cdot \nabla T  \over T},
\label{Phi}
\end{equation}
where $\tau_n$ is the effective relaxation time
of neutrons which is generally a complicated function
of the energy parameter $(\epsilon_n - \mu_n) / T$.
However, while solving the transport equation, we will rely on
the variational approach and choose the simplest trial function
consistent with a symmetry requirement
imposed on $\Phi_n$. This
corresponds to $\tau_n$ independent of energy.

Notice that the thermal conductivity
can be calculated exactly using the formalism developed by Sykes \&
Brooker (\cite{SB70}), H{\o}jgaard Jensen et al.\ (\cite{HJeta68}),
Flowers \& Itoh (\cite{FI79}),
and Anderson et al.\ (\cite{Aetal87}).
In Sect.\ \ref{s27} we discuss the corrections to our results
which follow from the exact theory.

\subsection{Relaxation time}
\label{s22}
Inserting $F_n$ from Eqs.\ (\ref{Fn}) and (\ref{Phi})
into (\ref{q}) we have
\begin{equation}
      \kappa_n = {\pi^2 T n_n \tau_n \over
       3 m^\ast_n},
\label{kappa}
\end{equation}
where $n_n$ and $m_n^\ast$ are the neutron
number density and effective mass, respectively.
The relaxation time $\tau_n$ has to be determined
from Eq.\ (\ref{kin}).

Following the standard procedure we linearize the
kinetic equation with respect to the small nonequilibrium
correction $\Phi_n$ to the Fermi--Dirac distribution.
The linearized left--hand side of Eq.\ (\ref{kin}) is
\begin{equation}
      {\vec v}_n \cdot \nabla F_n \approx {\vec v}_n \cdot \nabla
      f_n = {\partial f_n \over \partial T} \, \,
      {\vec v}_n \cdot \nabla T.
\label{apprleft}
\end{equation}
Inserting Eqs.\ (\ref{Fp}) and (\ref{Fn})
into (\ref{Igen}), we arrive at the linearized collision integrals
\begin{eqnarray}
     I_{12} & = & {\tau_n \, V^3 \nabla T
        \over (2 \pi)^9 \, (1+\delta_{12}) \, T^2} \, \int \!  \int \! \int
        \dd {\vec k}'_1 \, \dd {\vec k}_2 \,
        \dd {\vec k}'_2 \, f_1 \, f_2
\nonumber \\
        && \times \, (1-f'_1)  \, (1-f'_2)
        \sum_{\sigma'_1 \sigma_2 \sigma'_2}
        {\cal P}_{12} \, {\vec V}_{12},
\label{I_linear}
\end{eqnarray}
where $ {\vec V}_{nn} =  (\epsilon_1- \mu_n) \, {\vec v}_1 +
       (\epsilon_2-\mu_n) \, {\vec v}_2  -
       (\epsilon'_1- \mu_n) \, {\vec v}'_1  -
       (\epsilon'_2- \mu_n) \, {\vec v}'_2   $ and
       $ {\vec V}_{np}=
       (\epsilon_1-\mu_n) \, {\vec v}_1 - (\epsilon'_1- \mu_n) \,
       {\vec v}'_1 $.

Let us multiply the linearized Boltzmann equation
by ${\rm d}{\vec k}_n \: (2 \pi)^{-3} \:
(\epsilon_n - \mu_n) \, {\vec v}_n$,
integrate over d${\vec k}_n$ and sum over
the neutron spin states $\sigma_n$. From
the left--hand side of the Boltzmann equation
we obtain
\begin{equation}
       C_n =  \sum_{\sigma_n} \, \int {\dd {\vec k}_n \over (2 \pi)^3}
       \, (\epsilon_n-\mu_n) \, v^2_n \,{\partial f_n
       \over \partial T} = {\pi^2 T n_n \over m^\ast_n}.
\label{leftpart}
\end{equation}
On the right--hand side, we have $\pi^2 \tau_n T n_n
(\nu_{nn} + \nu_{np}) / m^*_n$, where we have introduced
the effective $nn$ and $np$ collision frequencies
$\nu_{nn}$ and $\nu_{np}$:
\begin{eqnarray}
     \nu_{12} &=& { m_n^\ast \, V^3 \over (2 \pi)^{12} \pi^2 T^3 n_n
        (1+\delta_{12})}
        \, \int \! \int \! \int \! \int
        \dd {\vec k}_1 \, \dd {\vec k}'_1 \,
\nonumber \\
        & & \times \;
        \dd {\vec k}_2 \, \dd {\vec k}'_2 \,
        \sum_{\sigma_1 \sigma'_1
        \sigma_2 \sigma'_2} {\cal P}_{12} \, f_1 \, f_2
\nonumber \\
        & & \times \;  
        \, (1-f'_1) \, (1-f'_2) \,
        (\epsilon_1-\mu_n) \, {\vec V}_{12} \cdot {\vec v}_1.
\label{nu_general}
\end{eqnarray}
Now the relaxation time is:
\begin{equation}
      \tau_n ={ 1 \over \nu_{nn} + \nu_{np}}.
\label{tau}
\end{equation}

The differential transition rate
summed over initial and final spin states can be written as:
\begin{eqnarray}
        \sum_{\sigma_1 \sigma'_1
        \sigma_2 \sigma'_2} {\cal P}_{12} & = &
        4 \, { (2 \pi)^4 \over \hbar \, V^3} \,
        \delta(\epsilon_1 + \epsilon_2 -
        \epsilon'_1 - \epsilon'_2) 
\nonumber \\
      && \times \;  \delta({\vec k}_1 + {\vec k}_2 -
        {\vec k}'_1 - {\vec k}'_2) \, {\cal Q}_{\rm 12},
\label{TransitionRate}
\end{eqnarray}
where ${\cal Q}_{12}= (1/4) \sum_{\rm spins} |M_{12}|^2$
is the squared matrix element
averaged over initial and summed over final spin states.
Since no specific spin orientations are involved,
${\cal Q}_{12}$ can depend
only on the scalar combinations of the momentum transfers
${\vec q} = {\vec k}'_2 - {\vec k}_2$ and
${\vec q}' = {\vec k}'_2 - {\vec k}_1$,
that is on $q$, $q'$ and $({\vec q} \cdot {\vec q}')$.
In our case, the scalar product $({\vec q} \cdot {\vec q}')$ is zero
(see Sect.\ \ref{s23}).
Therefore, ${\cal Q}_{12}$ is a function of $q$ and $q'$ alone.

\subsection{Energy--angular decomposition}
\label{s23}
Since neutrons and protons are strongly degenerate,
we can use the conventional energy--angular decomposition
while evaluating the collision frequencies $\nu_{12}$.
The main contribution
to the integrals comes from narrow
vicinities of $k$ near $k=k_{\rm F}$ in which 
$\hbar v_{\rm F} |k - k_{\rm F}| \la T$
(in this case $k_{\rm F}$ and $v_{\rm F}$ are a particle Fermi
wavenumber and velocity).
In these regions, we can introduce convenient
momentum variable  $\xi=\hbar v_{\rm F}(k-k_{\rm F})$ with
${\rm d}\xi = {\rm d}\epsilon$. Then we can replace integration
over $k$ by that over $\xi$ extending the integration limits
to $ \pm \infty$, since the domains of integration outside the
indicated vicinities give exponentially small contributions.
Thus $\hbar^2 {\rm d} {\vec k} =
k_{\rm F} m^\ast \, {\rm d} \xi \, {\rm d} \Omega$,
where d$\Omega$ is a solid angle element in the direction of
{\vec k}. Moreover,
we can place the particle wave vectors {\vec k}
at the appropriate Fermi surfaces
in all smooth functions of the momenta. 
Consequently, the expressions for $\nu_{12}$
are decomposed into sums of products
containing decoupled integrations over angles d$\Omega$ and
momentum (energy) variables d$\xi$ of all colliding particles.
Placing all the particle wave vectors on their
Fermi surfaces, one can easily verify that the momentum transfers
{\vec q} and ${\vec q'}$ are perpendicular, as already
mentioned in Sect.\ \ref{s22}. Using the well-known
symmetry properties of the energy integrals
and the relations which come from momentum conservation
for the particles at their Fermi surfaces in the
angular integrals, after standard transformations
(e.g., Ziman, \cite{Z60}) we obtain
\begin{eqnarray}
      \nu_{12} &  = & {3 m_1^\ast k_{\rm F1} m_2^{\ast 2} k_{\rm F2}^2
      \over 2^6 \pi^8 \, T^3 \, \hbar^7 
      (1+\delta_{12})} 
\nonumber \\
      & & \times \; \int \! \int \! \int \! \int_{-\infty}^{+\infty}
      \dd \xi_1 \, \dd \xi'_1 \,  \dd \xi_2  \,  \dd \xi'_2 \,
      f_1 \, f_2 
\nonumber \\
      & & \times \; (1-f'_1) \, (1-f'_2) \,
      \delta(\epsilon_1+ \epsilon_2-\epsilon'_1- \epsilon'_2) 
\nonumber \\
      & & \times \; \int \! \int \! \int \! \int \dd \Omega_1 \,
      \dd \Omega'_1 \,  \dd \Omega_2  \,  \dd \Omega'_2 
      \delta({\vec k}_1 + {\vec k}_2 - {\vec k}'_1 - {\vec k}'_2) 
\nonumber \\
      & & \times \;
      {\cal Q}_{12}(q,q') \, (\xi_1^2 - \xi_1 \xi'_1 A_{12}),
\label{nu_through_Q}
\end{eqnarray}
where $A_{nn}=3-(q^2+q^{\prime 2})/k_{\rm Fn}^2$,
$A_{np}=1- q^2/(2 k_{\rm Fn}^2)$.
These expressions contain two energy integrals $J_1$ and $J_2$
of the form
\begin{eqnarray}
      J_\alpha &=&  
      \int \! \int \! \int \! \int_{-\infty}^{+\infty} 
      \dd \xi_1 \, \dd \xi'_1 \, \dd \xi_2 \, \dd \xi'_2 \, 
      \delta(\epsilon_1+ \epsilon_2-\epsilon'_1- \epsilon'_2) 
\nonumber \\
      & & \times \; f_1 \, f_2
      \, (1-f'_1) \, (1-f'_2) \,
      (\epsilon_1-\mu_n) \, B_\alpha,
\label{J's}
\end{eqnarray}
with $B_{1}=\epsilon_1-\mu_1$ and
     $B_{2}=\epsilon'_1-\mu_1$.
They may be 
evaluated analytically:
\begin{equation}
      J_{1} = 3 \, J_{2} =
      {2 \over 5} \pi^4 T^5.
\label{JE}
\end{equation}
Thus we arrive at the equations which contain the angular integration
\begin{eqnarray}
      \nu_{12} &=& {16 \pi^5 \, m_1^\ast m_2^{\ast 2}
      \, k_{\rm F2}^2 \, T^2
      \over 5 \, k_{\rm F1} (2 \pi)^9 \, \hbar^7 (1+ \delta_{12})} \,
      \int \! \int \! \int \! \int \dd \Omega_1 \,
      \dd \Omega'_1 \,  \dd \Omega_2 \, \dd \Omega'_2 \,
\nonumber \\
      && \times \; \delta({\vec k}_1 + {\vec k}_2 -
      {\vec k}'_1 - {\vec k}'_2)
      \; {\cal Q}_{12}(q,q') \, C_{12},
\label{nu_through_Q1}
\end{eqnarray}
where $C_{nn}= q^2 + q'^2$ and $C_{np}= 2 k^2_{\rm Fn} + (q^2/ 2)$.

\subsection{Angular integration}
\label{s24}
It is clear that the integrands depend only on the relative
orientations of the particle momenta. Then we can
immediately integrate over the orientations
of the momentum of the first particle before the scattering,
and over the azimuthal angle which specifies the momentum of the second
colliding particle with respect to the first one.
This yields an overall factor $8 \pi^2$.
Let $\theta$ be an angle between the momenta of colliding particles
before the scattering,
and $\theta_1$, $\theta_2$, $\phi_1$, $\phi_2$
be the polar and azimuthal angles of the particles after the scattering
in a coordinate frame with $z$ axis along ${\vec k}_1$ and
$x$ axis placed in a (${\vec k}_1 \, {\vec k}_2$)--plane.
Accordingly, 
\begin{equation}
     \sin\theta_1 \, \dd \theta_1 = {q \, \dd q \over k^2_{\rm F1}},
     \quad \quad
     \sin\theta_2 \, \dd \theta_2 = {q' \, \dd q'
     \over k_{\rm F1} k_{\rm F2}},
\label{angle}
\end{equation}
and the integration over $\theta$, $\phi_1$, $\phi_2$ is
performed with aid of the $\delta$--function:
\begin{eqnarray}
    && {k_{\rm F2} \over k_{\rm F1}} \, \int \! \int \! \int \sin\theta \,
     \dd \theta \, \dd \phi_1 \, \dd \phi_2 \,
     \delta ({\vec k}_1+{\vec k}_2-
     {\vec k}'_1-{\vec k}'_2) 
\nonumber \\
    &&  =
     {4 \over q q' \sqrt{A^2 - q^2}},
\label{delk}
\end{eqnarray}
where
\begin{equation}
     A^2 = {1 \over q'^2} \,\left[4 k^2_{\rm F1} k^2_{\rm F2} -
     (k^2_{\rm F1}+k^2_{\rm F2}-q'^2)^2 \right],
\label{alit}
\end{equation}
$q \leq A$,
$| k_{\rm F1}-k_{\rm F2} | \leq  q' \leq (k_{\rm F1}+k_{\rm F2})$.

Inserting (\ref{delk}) into (\ref{nu_through_Q1})  we
can express $\nu_{nn}$ and $\nu_{np}$ in a unified manner
\begin{equation}
       \nu_{12} = {64 m_1^\ast m_2^{\ast 2} k_{\rm B}^2 T^2
       \over 5 m_N^2 \hbar^3 }
       \, S_{12}.
\label{nu_through_S}
\end{equation}
Here, we introduce the convenient quantities of dimension
of a cross section (cm$^2$):
$S_{nn} \equiv S_{n2}$ and $S_{np}  =  S_{p1} + S_{p2}$:
\begin{eqnarray}
       S_{n1} & = & {m_N^2 \over 16 \pi^2 \hbar^4}
       \int^1_0 \dd x' \int^{\sqrt{1 - x^{\prime 2}}}_0
       \dd x \, { 1 \over
       \sqrt{1 - x^2 - x^{\prime 2}}} \, {\cal Q}_{nn},
\nonumber \\
\nonumber \\
       S_{n2} & = & {m_N^2 \over 16 \pi^2 \hbar^4}
       \int^1_0 \dd x' \int^{\sqrt{1 - x^{\prime 2}}}_0
       \dd x \, {x^2+x^{\prime 2} \over
       \sqrt{1 - x^2 - x^{\prime 2}}} \, {\cal Q}_{nn},
\nonumber \\
\nonumber \\
       S_{p1} & = & {m_N^2 \over 16 \pi^2 \hbar^4}
       \int^{0.5+x_0}_{0.5-x_0}
       \dd x' \int^{a(x')}_0 \dd x \,
       {1  \over \sqrt{a^2-x^2}} \, {\cal Q}_{np},
\nonumber \\
\nonumber \\
       S_{p2} & = & {m_N^2 \over 16 \pi^2 \hbar^4}
       \int^{0.5+x_0}_{0.5-x_0}
       \dd x' \int^{a(x')}_0 \dd x \,
       { x^2  \over \sqrt{a^2-x^2}} \, {\cal Q}_{np}.
\label{Snp}
\end{eqnarray}
In this case, $x = q /(2 k_{\rm Fn})$, $x' = q' / (2 k_{\rm Fn})$,
$x_0 = k_{\rm Fp} /( 2 k_{\rm Fn})$,
$a(x') = \sqrt{x^2_0 - (0.25 + x^2_0 - x'^2 )^2} /x'$,
$m_N$ is a bare nucleon mass (we neglect the small difference
between the bare neutron and proton masses).
The quantity $S_{n1}$ is not needed here 
but will be required in Sect.\ \ref{s3}.

Thus we have expressed $\nu_{12}$
through the 2D integral of the squared matrix element
${\cal Q}_{12}$ over the particle momenta transfers $q$ and $q'$.
This integration can be performed numerically.
Note that for $nn$ collisions $q$ varies from 0 to 2$k_{\rm Fn}$
and scattering at arbitrary angles contributes to
the integration. By contrast, in the $np$ case $a(x')$ does not exceed
2$x_0$ that is scattering angles larger than 2 ${\rm arcsin}
(k_{\rm Fp}/k_{\rm Fn})$ are forbidden. We recall that, typically,
$k_{\rm Fp} \ll k_{\rm Fn}$, in neutron star cores.

At this stage it is convenient
to introduce the scattering cross section.

\subsection{Scattering cross section}
\label{s25}
The squared matrix element
${\cal Q}_{12}$ determines the
scattering cross section. The nucleon--nucleon cross section
is more convenient
for practical calculations
since it has been the subject of some studies.
To translate the cross section into the squared matrix element one should be
careful about kinematics. We are looking for a matrix element
between the quasiparticle states
with certain momenta. Let us consider an idealized
case, in which we can approximate nucleon quasiparticles by bare
nucleons.
The simplest choice would
be then to equate this matrix element to that for scattering of nucleons
with the same momenta in vacuum. This proceeds as follows. Let us
specify, for our imaginary bare nucleons, the center-of-mass (cm)
reference frame in which the total momentum
of a colliding pair is zero,
and the laboratory (lab) reference frame in which
the initial momentum of the second particle
is zero. Using standard quantum mechanics we find the familiar
relations:
\begin{eqnarray}
     {{\rm d} \sigma_{12} \over {\rm d} \Omega_{\rm cm}} \,
     (\epsilon_{\rm lab}, \theta_{\rm cm}) &=&
     {m_N^2 \over 16 \pi^2 \hbar^4} \, {\cal Q}_{12},
\label{BareCrossSection} \\
     {\rm cos} \theta_{\rm cm} &=& {q'^2 - q^2 \over q'^2 + q^2},
\label{thetaCM} \\
     \epsilon_{\rm lab} & = & {\hbar^2 \over 2 m_N} \, (q^2+q'^2).
\label{BareELAB}
\end{eqnarray}
In this representation, which 
is popular
in the current
literature, the cross section is a function of
scattering angle in the
center-of-mass frame and
collision energy in the laboratory frame.

However, there exists a more involved approximation based on
recent microscopic calculations of $N\!N$ 
scattering cross sections in medium
(see also Sect.\ \ref{s26}). To restore the squared matrix element
in this case, we again consider the two reference frames but
this time
for quasiparticles. The quasiparticle energy and
momentum transformations which accompany a Galilean transformation
from  one frame 
to another may be found in Sect.\ 1.1.3 of
Baym \& Pethick (\cite{BP91}). 
Although the scattering angle in the center-of-mass
reference frame is still given by Eq.\ (\ref{thetaCM}),
other expressions change:
\begin{eqnarray}
     {{\rm d} \sigma_{12} \over {\rm d} \Omega_{\rm cm}} &=&
            { (m^\ast_1 m^\ast_2)^2 (q^2+q'^2)^{3/2}\, {\cal Q}_{12}
             \over
            8 \pi^2 \hbar^4 }
\nonumber \\
           & & \times \; \left[(k^2_{\rm F1} m_2^\ast - k^2_{\rm F2} m_1^\ast)
     \left(m^\ast_2 - m^\ast_1 \right)
     \right.
\nonumber \\      
      && \left.  + \; 
       m_1^\ast m_2^\ast
     (q^2+q'^2) \right]^{-1/2}
\nonumber \\
        & & \times \;
         \left[(q^2+q'^2)(m^\ast_1+m^\ast_2) \right.
\nonumber \\
     && \left.  + \;
     (k^2_{\rm F1}-k^2_{\rm F2})(m^\ast_2-m^\ast_1) \right]^{-1},
\label{MedCross} \\
     \epsilon_{\rm lab} &=& {\hbar^2 \over 2 m_N} (q^2+q'^2) +
           {p^2_{\rm F1} \over 2 m_N} \left({m_N \over m^\ast_1} - 1 \right)
\nonumber \\
    &&  + \;
           {p^2_{\rm F2} \over 2 m_N} \left({m_N \over m^\ast_2} - 1 \right).
\label{MedELAB}
\end{eqnarray}
If the effective masses of colliding quasiparticles coincide,
Eq.\ (\ref{MedCross}) reduces to a much simpler
Eq.\ (\ref{BareCrossSection}), with
$ m^{\ast}$ in place of $m_N$.

\subsection{Calculation of $S_\alpha$}
\label{s26}
For the collision energies up to 550 MeV we are
interested in, the quark structure of matter is not yet pronounced,
and the bare $N\!N$ interaction can be described in terms
of various virtual--meson--exchange processes.

Very good agreement with experimental data on the elastic
$N\!N$ scattering
at energies up to 350 MeV is achieved using the
Bonn potential model (Machleidt et al.\ \cite{Metal87}). 
It includes all possible
diagrams which describe the exchange of $\pi$, $\rho$, $\omega$, $\eta$,
and $\delta$ mesons as well as the exchange of $2\pi$, $3\pi$,
$\pi + \rho$ mesons and other combinations of mesons with total
energy of any process up to 1 GeV. The maximum
collision energy of 350 MeV
is mentioned because the model does not
include the inelastic channels of $\pi$ meson creation
allowed at higher energies. We do not consider matter with
pion condensates, and, therefore, we do not allow for
these inelastic channels.

The bare $N\!N$ potential is known to be attractive
at large distances and repulsive at short ones.
The short--range repulsive core is extremely
strong, and the Born approximation fails to describe
the $N\!N$ collisions of interest.
Thus we use the exact transition probabilities
obtained with the Bonn potential.

The differential cross sections for free nucleons have been compiled
from tables presented by Li \& Machleidt (\cite{LM93,LM94}).
These authors consider {\it in-medium}
effects but their calculations are extended to the limit of zero
medium density which is the in-vacuum case. The latter cross sections
are computed for laboratory energies up to 300 MeV, and the results
are in nice agreement with experimental data. For
convenience, we prefer to use these theoretical data
rather than the experimental ones.
The cross sections of the $nn$
collisions at higher
energies have been set equal to their
values at 300 MeV since, according to the experimental data,
the $nn$ cross sections are almost energy--independent in
this range.

The treatment of the many-body effects is complicated.
For instance, Li \& Machleidt (\cite{LM93,LM94})
performed calculations for symmetric
nuclear matter as they were interested in heavy--ion collisions.
The detailed calculations for arbitrary asymmetry of matter,
in particular, for strongly asymmetric matter in neutron
star cores have not yet been done,
to our knowledge.
The many-body effects turn out to be rather significant;
they mainly reduce the scattering cross sections by a factor of 2--6.
Although the in-medium effects can obviously be
different in the symmetric and asymmetric matter,
we make use of the available results to illustrate possible
influence of medium on the transport properties.

We have calculated
the integrals (\ref{Snp}) using  both in-medium and in-vacuum sets
of the scattering cross sections
obtained with the Bonn model
for the values of $k_{\rm Fn}$
from 1.1 to 2.6 fm$^{-1}$ and for the values of $k_{\rm Fp}$ from 0.3 to
1.2 fm$^{-1}$ with the maximum ratio of $k_{\rm Fp}$ to $k_{\rm Fn}$
equal to 0.7. This grid certainly covers the range
of possible variation of $k_{\rm Fn}$ and $k_{\rm Fp}$ for any
realistic equation of state of matter at densities
$0.5 \rho_0 \la \rho \la 3 \rho_0$.
While restoring the squared matrix element from the in-medium
cross sections we set $m_n^\ast = m_p^\ast = 0.8 \, m_N$
in Eqs.\ (\ref{MedCross}) and (\ref{MedELAB}).

The numerical results are fitted by simple analytic
functions and written in the form
\begin{equation}
    S_\alpha = S_\alpha^{(0)} K_\alpha,
\label{SK}
\end{equation}
where $S_\alpha^{(0)}$ corresponds to scattering of bare
particles, and $K_\alpha$ describes the in-medium
effects. In all these fits, $k_{\rm Fn}$ and
$k_{\rm Fp}$ are expressed in fm$^{-1}$.

%

The fit expressions for the in-vacuum integrals are
\begin{eqnarray}
  S_{n1}^{(0)} & = &  {14.57 \over k_{\rm Fn}^{1.5}} \times
        {1 - 0.0788 \, k_{\rm Fn} + 0.0883 \, k_{\rm Fn}^2
        \over 1 - 0.1114 \, k_{\rm Fn} }~~{\rm mb},
\nonumber \\
 S_{n2}^{(0)} & = &  {7.880 \over k_{\rm Fn}^2} \times
        {1 - 0.2241 \, k_{\rm Fn} + 0.2006 \, k_{\rm Fn}^2
        \over 1 - 0.1742 \, k_{\rm Fn} }~~{\rm mb},
\nonumber \\
  S_{p1}^{(0)} & = &
         {0.8007 \, k_{\rm Fp} \over k_{\rm Fn}^2}
       \, (1+ 31.28 \, k_{\rm Fp} 
-  0.0004285 \, k_{\rm Fp}^2
\nonumber \\
       & &
        + \;  26.85 \, k_{\rm Fn} + 0.08012 \, k_{\rm Fn}^2 )
         \;
        ( 1 - 0.5898 \, k_{\rm Fn} 
\nonumber \\
   &&  + \;  0.2368 \, k_{\rm Fn}^2
+ 0.5838 \, k_{\rm Fp}^2 +
        0.884 k_{\rm Fn} k_{\rm Fp})^{-1}~{\rm mb},
\nonumber \\
  S_{p2}^{(0)} & = &
        {0.3830 \, k_{\rm Fp}^4 \over k_{\rm Fn}^{5.5}}
        (1+ 102.0 \, k_{\rm Fp}
        +53.91 \, k_{\rm Fn}) 
\nonumber \\
        && \times \;(
        1 - 0.7087 \, k_{\rm Fn} + 0.2537 \, k_{\rm Fn}^2
\nonumber \\
        &&  + \; 9.404 \, k_{\rm Fp}^2 - 1.589 \, 
          k_{\rm Fn} k_{\rm Fp})^{-1}~~{\rm mb}.
\label{S_0fit}
\end{eqnarray}
The mean fit errors of $S_{n1}^{(0)}$ and $S_{n2}^{(0)}$
are about $\delta_{\rm rms} \approx 0.3\%$,
and the maximum errors $\delta_{\rm max} \approx 0.5\%$ take place at
$k_{\rm Fn} \approx 1.5~{\rm fm}^{-1}$. In the case of
$S_{p1}^{(0)}$, we have $\delta_{\rm rms} \approx 1.2 \%$, with
$\delta_{\rm max} \approx 2.9 \%$ at $k_{\rm Fn} \approx
1.1~{\rm fm}^{-1}$ and $k_{\rm Fp} \approx 0.3~{\rm fm}^{-1}$,
while in the case of
$S_{p2}^{(0)}$, $\delta_{\rm rms} \approx 3.4 \%$, with
$\delta_{\rm max} \approx 9.0 \%$ at $k_{\rm Fn} \approx
1.1~{\rm fm}^{-1}$ and $k_{\rm Fp} \approx 0.4~{\rm fm}^{-1}$.


The fits to the in-medium correction factors $K_\alpha$ are
\begin{eqnarray}
    K_{n1} &  =  & \left( {m_N \over m_n^\ast} \right)^2
           ( 0.4583+ 0.892 \, u^2-0.5497 \, u^3
\nonumber \\           
       & &  - \; 0.06205 \, k_{\rm Fp}
               +0.04022\, k_{\rm Fp}^2 
    + 0.2122 \, uk_{\rm Fp} ),
\nonumber \\
    K_{n2} &  = & \left( {m_N \over m_n^\ast} \right)^2 
           ( 0.4891+ 1.111 \, u^2-0.2283 \, u^3
\nonumber \\
           && + \; 0.01589 \, k_{\rm Fp}   -0.02099\, k_{\rm Fp}^2 
              + 0.2773 \, uk_{\rm Fp} ),
\nonumber \\
     K_{p1} & = & \left( {m_N \over m_p^\ast} \right)^2 
           ( 0.04377 + 1.100 \, u^2 +0.1180 \, u^3
\nonumber \\
           && + \; 0.1626 \, k_{\rm Fp}   + 0.3871 \, uk_{\rm Fp}
             - 0.2990 u^4 ),
\nonumber \\
     K_{p2}& = & \left( {m_N \over m^\ast_p} \right)^2 
           ( 0.0001313 + 1.248 \, u^2 +0.2403 \, u^3
\nonumber \\
           &&  + \; 0.3257 \, k_{\rm Fp}  + 0.5536 \, uk_{\rm Fp}
     - 0.3237 u^4
\nonumber \\
         &&  + \; 0.09786\, u^2 k_{\rm Fp} ).
\label{K_fit}
\end{eqnarray}
In the case of $K_{n1}$ we define $u=k_{\rm Fn}-1.665$;
the mean fit error is about $\delta_{\rm rms} \approx 1.3\%$,
and the maximum error $\delta_{\rm max} \approx 3.9\%$ takes place at
$k_{\rm Fn} \approx 1.1~{\rm fm}^{-1}$ and
$k_{\rm Fp} \approx 0.3~{\rm fm}^{-1}$.
For $K_{n2}$, we have $u=k_{\rm Fn}-1.556$,
$\delta_{\rm rms} \approx 1.9\%$,
$\delta_{\rm max} \approx 4.7\%$ at
$k_{\rm Fn} \approx 1.8~{\rm fm}^{-1}$ and
$k_{\rm Fp} \approx 1.2~{\rm fm}^{-1}$.
For $K_{p1}$, we have $u=k_{\rm Fn} - 2.126$,
$\delta_{\rm rms} \approx 2.4 \%$,
$\delta_{\rm max} \approx 7.1 \%$ at $k_{\rm Fn} \approx
1.6~{\rm fm}^{-1}$ and $k_{\rm Fp} \approx 1.1~{\rm fm}^{-1}$.
Finally, for $K_{p2}$, we have $u=k_{\rm Fn} - 2.116$,
$\delta_{\rm rms} \approx 3.9 \%$,
$\delta_{\rm max} \approx 8.5 \%$ at $k_{\rm Fn} \approx
2.1~{\rm fm}^{-1}$ and $k_{\rm Fp} \approx 0.5~{\rm fm}^{-1}$.

Unfortunately, the quantities
$K_\alpha$ depend also implicitly on
$m^\ast$, as the latter enters Eq.\ (\ref{MedELAB})
for $\epsilon_{\rm lab}$. For this reason, Eqs.\
(\ref{K_fit}) are strictly valid only if
$m^\ast_n=m^\ast_p=0.8 \, m_N$. We will use these quantities 
for illustration and present the final results in the form
which could be modified easily if more advanced in-medium data
are available.

\subsection{Comparison with exact solution}
\label{s27}
As pointed out in Sect.\ \ref{s21}, we have used
the variational
method while solving the kinetic equation. In principle,
it is possible to derive an exact thermal conductivity,
making use of the advanced theory
developed by Sykes \& Brooker (\cite{SB70}), and
H{\o}jgaard Jensen et al.\ (\cite{HJeta68}),
for one--component
Fermi--systems, and extended by Flowers \& Itoh (\cite{FI79}) and
Anderson et al.\ (\cite{Aetal87}) to multi--component systems.
In our case neutrons are the only heat carriers, and
the equation which determines the energy dependence
of the nonequilibrium term in the neutron distribution
can be reduced to that for a one--component system:
\begin{eqnarray}
   &&    x f(x) [1-f(x)] = {x^2 + \pi^2 \over 2} \,
                         f(x) [1-f(x)] \Psi(x)
\nonumber \\
                         & & + \; \lambda_K \, \int^{\infty}_{-\infty}
                         {\rm d} x_2 \, {x + x_2 \over 1 -
                         {\rm e}^{-x - x_2}} \, f(x) f(x_2) \Psi(x_2).
\label{SBeq}
\end{eqnarray}
In this case, $x = (\epsilon_n - \mu_n) / T$, $f(x)=1/({\rm e}^x+1)$, 
and $\Psi(x)$ is an unknown function
defined in the same fashion as in Baym \& Pethick (\cite{BP91}).
The parameter $\lambda_K$, originally introduced by Sykes \&
Brooker (\cite{SB70}),
contains information on $nn$ and $np$ scattering
and reads
\begin{eqnarray}
    \lambda_K & = & \left\{ 
                   \int {{\rm d} \Omega \, (1 + 2 \cos{\theta})
                   \over \cos{(\theta/2)}} {\cal Q}_{nn} (\theta, \phi)
                   \right. 
        +   8 \, {m^{\ast 2}_p \over m^{\ast 2}_n} \,
                   x_0
\nonumber \\  
   & & \times \, 
                  \int {\rm d} \Omega \,
                   [(1+ 2 \, x_0 \, \cos{\theta})^2 +
                   4 \, x^2_0 \, \sin^2{\theta} \, \cos{\phi}]         
\nonumber \\
    & & \times \,  \left.           
                   {{\cal Q}_{np} (\theta, \phi_{\rm cm})
                    \over (1+4 \, x_0 \, \cos{\theta} + 4 \, x^2_0)^{3/2}} 
                   \right\} \; \left[
                   \int {{\rm d} \Omega \, {\cal Q}_{nn} (\theta, \phi)
                   \over \cos{(\theta/2)}} \right. 
\nonumber \\
    & &  +  \,  \left.
                   8 \, {m^{\ast 2}_p \over m^{\ast 2}_n} \,
                   x_0 \int {{\rm d} \Omega \, {\cal Q}_{np}
                   (\theta, \phi_{\rm cm})
                   \over (1+ 4 \, x_0 \cos{\theta} + 4\, x^2_0)^{1/2}}
                   \right]^{-1},
\label{lamK}
\end{eqnarray}
where $\theta$ and $\phi$ are Abrikosov--Khalatnikov angles, $x_0=
k_{\rm Fp}/(2k_{\rm Fn})$, and the integrations
are carried out over the
full  solid angle (${\rm d} \Omega = \sin{\theta} \,
{\rm d} \theta \, {\rm d} \phi $). In the case of $nn$ collisions,
$\phi$ is the scattering angle in the center-of-mass
reference frame, while for $np$ collisions we obtain
\begin{equation}
          \cos{\phi_{\rm cm}} = {(1-4 \, x^2_0)^2 +
          16 \, x^2_0 \, \sin^2{\theta} \, \cos{\phi}
          \over (1+ 4 \, x^2_0)^2 - 16 \, x^2_0 \,  \cos^2{\theta}}.
\label{fi_cm}
\end{equation}
The angle $\theta$ is connected, in both cases, with the collision energy
in laboratory reference frame (in-vacuum kinematics):
\begin{equation}
          \epsilon_{\rm lab} = {\hbar^2 \over 2 m_N} \,
                   (k^2_{\rm F1} + k^2_{\rm F2} - 2 k_{\rm F1} \, k_{\rm F2}
                   \cos{\theta}).
\label{teta-E}
\end{equation}

The exact thermal conductivity of neutrons
$\kappa_{\rm exact}$ is related to
the variational thermal conductivity $\kappa_{\rm var}$ 
through the correction factor $C$:
\begin{equation}
   \kappa_{\rm exact}=\kappa_{\rm var} \, C(\lambda_K),  \quad \quad
   C(\lambda_K)= {12 \over 5} H(\lambda_K),
\label{correction}
\end{equation}
where
\begin{eqnarray}
  &&   H(\lambda_K)  =  { 3 - \lambda_K \over 4}
\nonumber \\
      && \times \;
        \sum_{n=0}^\infty
        { 4n+5 \over (n+1)(2n+3)[(n+1)(2n+3)-\lambda_K ]}
\label{H}
\end{eqnarray}
is the function introduced by Sykes \& Brooker (\cite{SB70}).
For the range of particle Fermi wave-numbers $k_{\rm Fn}$ and $k_{\rm Fp}$
considered in the present paper,  the parameter $\lambda_K$
determined from Eq.\ (\ref{lamK})
varies from 0.79 to 0.97. Corresponding values of $C(\lambda_K)$ lie
between 1.20 and 1.22. It seems reasonable, therefore, to adopt the constant
correction factor $C=1.2$ to the values of $\kappa_{\rm var}$.

\subsection{Comparison with other works}
\label{s28}

Let us compare our results with the results of several previous
works at lower densities.
First of all, we mention two papers by
Wambach, Ainsworth \& Pines
(\cite{Wetal93}) (hereafter WAP) and
by Sedrakian et al.\ (\cite{Setal94}).
WAP considered pure neutron matter at densities
about $n_0 = 0.16$ fm$^{-3}$ using Landau theory.
They calculated momentum dependent Landau parameters,
quasiparticle transition amplitudes
and transport coefficients (thermal conductivity,
viscosity, and spin diffusion coefficient).
As follows from their Fig.\ 10, panel 4
(where the quantity $\kappa T/c$
should be expressed in MeV/fm$^2$ rather than
in K/fm$^2$ as printed), the thermal conductivity is
$\kappa^{\rm WAP} \approx 2 \times 10^{21}$ ergs s$^{-1}$
cm$^{-1}$ K$^{-1}$,
if we take $n_n = n_0$ and $T=$ 2 MeV.
(This particular temperature is chosen
for comparison with the results of Sedrakian et al. (\cite{Setal94});
other authors neglected temperature effect on
the transition probabilities and, consequently, found the standard
$T^{-1}$ dependence of the thermal conductivity.)
The value of the Landau parameter $F^s_1$ at zero momentum
transfer from WAP's Fig.\ 7
yields the effective mass $m^\ast_n \approx 0.9 m_{\rm N}$ at
$n_n = n_0$. Using our in--vacuum formula
with the same $m_n^\ast$ and omitting
$np$ collisions, we obtain
$\kappa \approx 2.9 \times 10^{20}$ ergs s$^{-1}$ cm$^{-1}$ K$^{-1}$.
With the medium effects included
by using the results
of  Sect.\ \ref{s26}
we get about $4.7 \times 10^{20}$ ergs s$^{-1}$ cm$^{-1}$ K$^{-1}$.
In this case, we have taken into account
rather weak dependence 
of the in-medium correction $K_n$, Eq.\ (\ref{K_fit}),
on $k_{\rm Fp}$.

The calculation by Sedrakian et al.\ (\cite{Setal94}),
based on a thermodynamic $T$--matrix approach, yields
$\kappa^{\rm S} \approx 2.6 \times 10^{20}$
ergs s$^{-1}$ cm$^{-1}$ K$^{-1}$
which agrees surprisingly well with our in--vacuum result
for the same $T$=2 MeV and $n_n=n_0$.
However, the treatment of Sedrakian et al.\ (\cite{Setal94}) is limited
to rather high temperatures $T > 1$ MeV at $\rho \sim \rho_0$.
At lower temperatures, the authors observe
a divergence of their $nn$ scattering cross--sections
indicating
the onset of neutron superfluidity. It is worth noting also, that
Fig.\ 2b of Sedrakian et al. (\cite{Setal94}) 
reproduces inaccurately
the thermal conductivity obtained by WAP. For instance, it gives
$2.7 \times 10^{20}$ ergs s$^{-1}$ cm$^{-1}$ K$^{-1}$, for
$T$ and $n_n$ under discussion, that is 7.5 times smaller
than the actual value quoted above.

Finally, our thermal conductivity based on in--vacuum
cross--sections can be directly compared with
the neutron thermal conductivity obtained by Flowers \& Itoh
(\cite{FI79}). Their results are valid
for a model of dense asymmetric nuclear matter proposed
by Baym et al.\ (\cite{BBP71}). From Fig.\ 1 of Flowers \& Itoh 
(\cite{FI79}) we have
$\kappa_n^{\rm FI} \approx 10^{23}$ ergs s$^{-1}$ cm$^{-1}$ K$^{-1}$
at $\rho \approx 2.7 \times 10^{14}$ g cm$^{-3}$ and
$T=10^8$ K. Corresponding particle number densities
are (Baym et al., \cite{BBP71}) $n_n \approx 0.153$ fm$^{-3}$, and
$n_p \approx 0.006$ fm$^{-3}$.
The nucleon effective masses were taken equal to
their bare masses.
In this case, our formulae yield
the value $\kappa_n \approx 1.8 \times 10^{22}$
ergs s$^{-1}$ cm$^{-1}$ K$^{-1}$,
which is more than 5 times smaller.
Note that Flowers \& Itoh (\cite{FI79}) reported good agreement
between their results
obtained in two ways: from the measured vacuum phase shifts and
in the spirit of Landau theory. 
A qualitative agreement of their results with the Landau--theory
results by WAP is clear from the above discussion.
However the accuracy of their vacuum phase shift approach
is questionable since $\kappa^{\rm FI}$ disagrees noticeably with our
in--vacuum thermal conductivity. The latter is virtually exact
(see Sect.\ \ref{s26}) and should be the same
as obtained from the measured vacuum phase shifts.

Summarizing we stress a large scatter of results obtained
using various approaches. In this respect,
a reliable calculation of the
Landau parameters and momentum dependent quasiparticle
amplitudes in the density range
(2 -- 3) $\rho_0$, at least
for pure neutron and symmetrical nuclear matter, is highly
desirable. It is especially true in view of relatively
large discrepancies at low densities between the calculations
which employ the in--vacuum cross--sections (like ours) and
those performed in the frames of the Landau theory.

\section{Effects of Superfluidity of Nucleons}
\label{s3}
The equations of Sect.\ \ref{s2}
give the
neutron thermal conductivity for non-superfluid
nucleons. Now we focus on
the effects of nucleon superfluidity.

It is well known (e.g., Tilley \& Tilley, \cite{TT90}) 
that the heat transport problem
in superfluid matter is complicated by the appearance
of convective counterflow of normal part of
fluid component in the presence of a
temperature gradient. Macroscopic counterflow
can carry heat producing
an effective thermal conductivity. A study
of this effective conductivity is a delicate task
which is outside the scope of the present work.
We will consider the effect of superfluidity
on the microscopic (diffusive) thermal conductivity.
This problem is also complicated and we will
adopt some model assumptions. Our consideration
is basically close to that used in the
studies of transport properties in superfluid $^3$He
by Bhattacharyya et al.\ (\cite{BPS77})
and Pethick et al.\ (\cite{PSB77}).  

It is generally believed (see, e.g., Takatsuka \& Tamagaki, \cite{TT93})
that the nucleon superfluidity
in the neutron star cores is of the BCS type.
The proton pairing occurs mainly in the singlet $^1S_0$
state. The neutron pairing can be of the singlet type only
at not too high densities, $\rho \la \rho_0$,
that is near the outer boundary of the neutron star core.
At higher densities, the $nn$
interaction in the
$^1S_0$ state becomes repulsive, and the singlet--state
neutron pairing disappears. However, the $nn$
interaction in the triplet $^3P_2$ state
is attractive, and the neutron
superfluidity at densities $\rho \ga \rho_0$
is most likely to occur in this state.
We consider the singlet--state proton
superfluidity, and either the singlet--state or
triplet--state neutron superfluidity.


Microscopically, the superfluidity leads to the appearance of an
energy gap $\delta$ in the superfluid quasiparticle dispersion
relation near
the Fermi surface ($ |k - k_{\rm F} | \ll k_{\rm F})$:
\begin{equation}
    \epsilon={\rm sgn}(\xi)\sqrt{\delta^2+\xi^2}~,
\label{supfdefs}
\end{equation}
where $\xi = \hbar v_{\rm F} (k-k_{\rm F}) $ 
is the normal-state quasiparticle energy measured
with respect to the chemical potential and
$v_{\rm F}$ is the normal-state quasiparticle  velocity at
the Fermi surface.
In the case of the singlet--state pairing, the energy
gap $\delta= \delta_0$ is isotropic, i.e., 
independent of orientation of the particle
momenta with respect to a quantization axis.
The temperature dependence of $\delta_0$
can be fitted as (e.g., Levenfish \& Yakovlev, \cite{LYheat})
\begin{eqnarray}
      y & = & {\delta_{\rm 0} (T) \over T} =
     \sqrt{1-\tau} \, \left(1.456 - {0.157 \over \sqrt{\tau}} 
      +   {1.764 \over \tau} \right),
\label{delta0}
\end{eqnarray}
where $\tau=T/T_{\rm c}$ and $T_{\rm c}$ is the critical temperature.

In contrast to the isotropic singlet--state pairing, the pairing in a
triplet--state produces
an anisotropic gap which depends on
orientation of nucleon momentum.
If the direction of the quantization axis were fixed
for some physical reason,
this would complicate the analysis because
the conductivity would be
anisotropic with respect to the quantization axis.
For the sake of simplicity, it is usually 
assumed  that
the orientation of the quantization axis
in the neutron star cores
is uncorrelated with rotational axis, temperature gradient,
or magnetic field.
Therefore, matter
of the neutron star
cores 
is treated as  
a collection of microscopic domains with arbitrary
orientations of the quantization axis.

Thus we assume that 
the diffusive thermal conductivity is isotropic even in the
presence of the triplet--state neutron pairing.
Under this assumption, we may use an effective dispersion
relation with an isotropic energy gap $\delta = \delta_1$
which greatly simplifies our analysis. It
would be inappropriate, though, to employ the same temperature
dependence of the gap as for the singlet--state pairing gap $\delta_0$,
Eq.\ (\ref{delta0}). For instance, this would
yield inaccurate gap parameters at $T \ll T_{\rm c}$.
We will take advantage of the detailed analysis
by Levenfish \& Yakovlev (\cite{LYdurca}),
Levenfish \& Yakovlev (\cite{LYheat}),
Yakovlev \&  Levenfish (\cite{YL95}), Yakovlev et al.\ (\cite{yls99}) of
the nucleon heat capacity and various neutrino--energy loss
rates in the neutron star cores for the case of triplet--state
pairing of the neutrons with zero projection
of the total angular momentum of
Cooper pairs onto the quantization axis ($m_J = 0$).
According to these studies, one can obtain qualitatively
accurate results using the isotropic gap
$\delta_1$ equal to the minimum value of the angle--dependent
gap on the Fermi surface. The temperature dependence
of this artificially isotropic gap can be fitted as 
(e.g., Yakovlev \& Levenfish, \cite{YL95}):
\begin{equation}
      y = {\delta_1 (T) \over T} =
     \sqrt{1-\tau} \, \left(0.7893 +
     {1.188 \over \tau} \right).
\label{delta1}
\end{equation}

Low--lying excited states of a nucleon superfluid correspond to the
presence of a dilute gas of superfluid quasiparticles. The time and
space
evolution of the quasiparticle distribution function is governed
by the transport equation. On the left-hand-side
(streaming terms) there appear two terms containing derivatives
of the superfluid gap but they cancel out. As a result,
the left-hand-side acquires
the standard form, identical to
that for normal quasiparticles.

However, the collision term is,
in general,  more complicated than that for normal nucleon
liquid. The scattering
amplitude for
quasiparticles in the superfluid is a linear
combination of the normal--state
amplitudes. These effects were considered in detail 
by Bhattacharyya et al.\ (\cite{BPS77}) at $T\la T_c$ and 
by Pethick et al.\ (\cite{PSB77}) at $T \ll T_c$
for liquid $^3$He. 
The accurate expressions for the superfluid scattering amplitudes
are complicated. 
In view of the large uncertainty in the amplitudes 
even in the case of non-superfluid
matter, we will keep the transition
probabilities the same as in the normal case. 
Thus we restrict ourselves to the most important
phase space effects stemming from the modification of the 
quasiparticle dispersion relations and resulting in a specific temperature
dependence of the collision frequencies. 
Our results at $T \la T_c$ and $T \ll T_c$
are in qualitative agreement with those of Bhattacharyya et al.\
(\cite{BPS77}) and Pethick et al.\ (\cite{PSB77}).

In order to calculate the diffusive thermal conductivity
in superfluid matter we adopt the same variational approach
which was used in Sect.\ \ref{s2} but with the modified dispersion
relations, Eq.\ (\ref{supfdefs}), containing energy gaps.
In particular, the deviation function is given by the
same Eq.\ (\ref{Phi}), where the velocity 
is now given by $v=v_{\rm F} | \xi | / \sqrt{\xi^2 + \delta^2}$.
Thus, the velocity of superfluid particles 
varies sharply near the Fermi surface 
and vanishes at $k=k_{\rm F}$. 

The temperature dependences of the collision integrals
can be obtained using the expressions for
the energy integrals, $J_\alpha$,
but with dispersion relations of
superfluid quasiparticles.
Notice that the combination of
Fermi--Dirac distributions in the collision integral remains
a sharp function of momenta which justifies our energy--angular
decomposition (Sect.\ \ref{s23}).
Then from Eq.\
(\ref{J's}) we obtain
\begin{equation}
     J_{\alpha} = J_{\alpha}^{(0)} \, {\cal R}_\alpha,
\label{Reduc}
\end{equation}
where $J_\alpha^{(0)}$ is an energy integral (\ref{JE})
in non-superfluid matter and
${\cal R}_\alpha$
describes the effect of nucleon superfluidity.
As we use the same ${\cal Q}_{12}$ as in the normal state, the
factor ${\cal R}_\alpha$ represents, in our approximation,
the overall effect of nucleon superfluidity on the collision
integral. Our analysis shows that ${\cal R}_\alpha <1$
in the presence of neutron and/or proton superfluidity.
In other words, the superfluidity always suppresses
the $nn$ and $np$ collision frequencies. There is no
discontinuity of ${\cal R}_\alpha$ when a superfluidity
is switched on with decreasing $T$. The suppression
becomes very large (${\cal R}_\alpha \ll 1$) for strong superfluidity
($T \gg T_c$).

Following Levenfish \& Yakovlev (\cite{LYdurca}) and
Yakovlev \&  Levenfish (\cite{YL95}) we introduce the notations
\begin{equation}
        x = {\xi \over T}, \quad
        y = {\delta \over T}, \quad
        z = {\epsilon \over T} = {\rm sgn}(x) \sqrt{x^2 + y^2}.
\label{ex-var}
\end{equation}
Then the correction factors for the $np$ collisions 
can be written as:
\begin{eqnarray}
        {\cal R}_{p1}(y_n,y_p) & \equiv &
        {5 \over 2 \pi^4} \int \! \int \! \int \! \int^{+\infty}_{-\infty}
        \dd x_n \dd x_p \dd x'_n
        \dd x'_p \, 
\nonumber \\
        && \times \; \delta (z_n+ z_p+z'_n+z'_p)
\nonumber \\
        & & \times \;  f(z_n) \, f(z_p)
        \, f(z'_n) \, f(z'_p) \, x_n^2,
\nonumber \\
        {\cal R}_{p2}(y_n,y_p) & \equiv &
        {15 \over 2 \pi^4} \int \int \int \int^\infty_{-\infty}
        \dd x_n \dd x_p \dd x'_n
        \dd x'_p \, 
\nonumber \\        
        && \times \; \delta (z_n+
        z_p+z'_n+z'_p) 
\nonumber \\
        & & \times \; f(z_n) \, f(z_p)
        \, f(z'_n) \, f(z'_p) \, (-x_n x'_n),
\label{R_p}
\end{eqnarray}
where $f(z)=({\rm e}^z+1)^{-1}=1-f(-z)$.
The reduction factors for the $nn$ collisions
are immediately expressed through ${\cal R}_{p \alpha}$:
\begin{equation}
       {\cal R}_{n\alpha}(y_n) = {\cal R}_{p \alpha}(y_n,y_n).
\label{R_n}
\end{equation}
The normalization factors in Eqs.\ (\ref{R_p}) 
are chosen to satisfy the condition
${\cal R}_{\alpha} = 1$ for vanishing gaps.
Furthermore, it is possible to analyze the asymptotic behaviour
of ${\cal R}_\alpha$ in the cases in which
the neutron and/or proton superfluidity is strong ($y \gg 1$).
This can be done in the standard manner
(cf.\ Levenfish \& Yakovlev, \cite{LYdurca},
Levenfish \& Yakovlev, \cite{LYheat},
Yakovlev \&  Levenfish, \cite{YL95}, Yakovlev et al., \cite{yls99})
Let us summarize the results.

If the neutron superfluidity is strong ($y_n \gg 1$),
the $nn$
collision frequency is affected by the factor
\begin{equation}
         {\cal R}_{n1}(y_n) \approx {\cal R}_{n2}(y_n) \approx
         \frac{15}{4 \pi^2} \, y^3_n \, {\rm e}^{ -2y_n}.
\label{Rnn_asy}
\end{equation}

In addition, we have calculated ${\cal R}_{n\alpha}(y_n)$
numerically and proposed analytic fits which
reproduce the numerical results and the above asymptotes:
%
%
\begin{eqnarray}
  {\cal R}_{n1}(y_n) & = & {2 \over 3} \,
     \left[0.9468 + \sqrt{(0.0532)^2 + 0.5346 \, y_n^2} \, \right]^3 
\nonumber \\
    && \times \;
     \exp \left[ 0.377 - \sqrt{ (0.377)^2+4 \, y_n^2 } \, \right] 
\nonumber \\
    & & + \; {1 \over 3} \, (1+ 1.351 \, y_n^2)^2 
\nonumber \\
    && \times \;  
     \exp \left[ 0.169 - \sqrt{ (0.169)^2+ 9 \, y_n^2 } \, \right],
\nonumber \\
%
%
  {\cal R}_{n2}(y_n) & = & {1 \over 2} \,
     \left[0.6242 + \sqrt{(0.3758)^2 + 0.07198 \, y_n^2} \, \right]^3 
\nonumber \\
     && \times \;
     \exp \left[ 3.6724 - \sqrt{ (3.6724)^2+4  \, y_n^2 } \, \right] 
\nonumber \\
    & & + \; {1 \over 2} \, (1+ 0.01211 \, y_n^2)^9 
\nonumber \\
    && \times \; 
     \exp \left[ 7.5351 - \sqrt{ (7.5351)^2+ 9 \, y_n^2 } \, \right],
\label{Rn_fit}
\end{eqnarray}
Maximum errors of these fits do not exceed 0.5\%.

Regarding $np$ collisions, we have calculated
${\cal R}_{p\alpha}(y_n,y_p)$ numerically for
the cases in which either neutron or proton
superfluidity is absent and fitted the numerical
results by the expressions:
%
%
\begin{eqnarray}
    R_{p1}(y_n,0) & = &
     \left[0.4459 + \sqrt{(0.5541)^2 + 0.03016 \, y_n^2} \, \right]^2 
\nonumber \\
     && \times \;
     \exp \left[ 2.1178 - \sqrt{ (2.1178)^2+ y_n^2 } \, \right],
\nonumber \\
%
%
    R_{p2}(y_n,0) & = &
     \left[0.801 + \sqrt{(0.199)^2 + 0.04645 \, y_n^2} \, \right]^2 \,
\nonumber \\
     && \times \;
     \exp \left[ 2.3569 - \sqrt{ (2.3569)^2+ y_n^2 } \, \right],
\nonumber \\
%
%
    {\cal R}_{p1}(0,y_p) & = & {1 \over 2} \,
     \left[0.3695 + \sqrt{(0.6305)^2 + 0.01064 \, y_p^2} \, \right] \,
\nonumber \\
     && \times \;
     \exp \left[ 2.4451 - \sqrt{ (2.4451)^2+ y_p^2 } \, \right]
\nonumber \\
     & & + \; {1 \over 2} \, (1+ 0.1917 \, y_p^2)^{1.4} \,
\nonumber \\
     &&  \times \;
     \exp \left[ 4.6627 - \sqrt{ (4.6627)^2+ 4 \, y_p^2 } \, \right],
\nonumber \\
%
%
    {\cal R}_{p2}(0,y_p) & = & 0.0436 \,
     \left[ \sqrt{(4.345)^2 + 19.55 \, y_p^2} -3.345 \, \right] \,
\nonumber \\
      && \times \;  \exp \left[ 2.0247 - \sqrt{ (2.0247)^2+ y_p^2 } \, \right] 
\nonumber \\
     &  & + \; 0.0654 \,
     \exp \left[ 8.992 - \sqrt{ (8.992)^2+ 1.5 \, y_p^2 } \, \right]
\nonumber \\
    & & + \, 0.891 
     \exp \left[ 9.627\! -\! \sqrt{ (9.627)^2+ 9  y_p^2 } \, \right].
\label{singleSF}
\end{eqnarray}
These fits reproduce numerical results
calculated in the range $y_n \le 30$
or $y_p \le 30$. The maximum fit error does not exceed
1\%, 2\%, 0.3\%, 1.4\%, respectively, and the mean
fit error is about twice lower.

The most difficult is the case of $np$ collisions
in matter in which neutrons and protons are superfluid at once.
In this case we have calculated ${\cal R}_{p1,2}(y_n, y_p)$
numerically on a dense grid of $y_n$ and $y_p$
($y_n \le 12$, $y_p \le 12$) and obtained analytic fits
to these results:
%
%
%
\begin{eqnarray}
      &&  {\cal R}_{p1} (y_n, y_p) =
       {\rm e}^{- u_+ - u_-} \, \left(0.7751
       + 0.4823 \, u_n + 0.1124 \, u_p  \right.
\nonumber \\
       & & + \; \left.  0.04991 \, u_n^2 +  0.08513 \, u_n u_p +
       0.01284 \, u_n^2 u_p   \right) 
\nonumber \\
        & & + \; {\rm e}^{- 2 u_+} \, \left(
       0.2249 \, +  0.3539 \, u_+ - 0.2189 \, u_-  \right.
\nonumber \\
        & & - \; \left.
        0.6069 \, u_n u_- + 0.7362 \, u_p u_+ \right),
\label{Rp1_fit}
\end{eqnarray}
where
$u_\beta = \sqrt{y_\beta^2 + (1.485)^2} - 1.485$,
$~\beta=~+,~-,~n,~p$,
$y_-={\rm min}(y_n,y_p)$ and $y_+={\rm max}(y_n,y_p)$.
The mean fit error is about 3\%, and the
maximum error of 14\% takes place at
$y_n \approx 10$ and $y_p \approx 12$,
where ${\cal R}_{p1}$ is exponentially small.

The fit to ${\cal R}_{p2}$ reads
%
%
\begin{eqnarray}
      && {\cal R}_{p2} (y_n, y_p) =
       {\rm e}^{- u_+ - u_-} \, \left(1.1032
       + 0.8645 \, u_n + 0.2042 \, u_p  \right.
\nonumber \\
       & & + \; \left.  0.07937 \, u_n^2 +  0.1451 \, u_n u_p +
       0.01333 \, u_n^2 u_p   \right) 
\nonumber \\
       & & + \;   {\rm e}^{- 2 u_+} \, \left(
       -0.1032 \, -  0.2340 \, u_+ + 0.06152 \, u_n u_+  \right.
\nonumber \\
       & & + \;   \left.
       0.7533 \, u_n u_- - 1.007 \, u_p u_+ \right),
\label{Rp2_fit}
\end{eqnarray}
where
$ u_\beta = \sqrt{y_\beta^2 + (1.761)^2} - 1.761$ with
$\beta=~+,~-,~n,~p$.
The mean fit error is about 3.6\%, and the
maximum error of 14\% again takes place at
$y_n \approx 10$ and $y_p \approx 12$ where ${\cal R}_{p2}$
is exponentially small. The accuracy of our fit formulae,
Eqs.\ (\ref{Rp1_fit}) and (\ref{Rp2_fit}), is quite
sufficient for practical applications. One can easily see
that these fit formulae are in good agreement with
Eqs.\ (\ref{Rn_fit}) at $y_n=y_p$ and with
Eqs.\ (\ref{singleSF}) either at $y_n=0$ or at $y_p=0$.

The neutron superfluidity affects also the left--hand side
of the kinetic equation given by the integral (\ref{leftpart}).
For a non-superfluid matter, we obtained $C_{n0}=k^3_{\rm Fn}
T /(3 m^\ast_n)$ (Sect.\ \ref{s22}).
In the presence of the neutron superfluidity, we can write
\begin{eqnarray}
       C_n (y_n) & = & C_{n 0} {\cal R}_C (y_n), 
\nonumber \\
       {\cal R}_C (y_n) & = & {6 \over \pi^2} \,
       \int_0^\infty {\rm d}x_n \: x_n^2 \,
       { {\rm e}^{z_n} \over ({\rm e}^{z_n}-1)^2}~.
\label{Cy}
\end{eqnarray}
In addition, the neutron superfluidity
affects the neutron heat flux in Eq.\ (\ref{q}). It turns out that
this effect is described by the same factor ${\cal R}_C$.
Consequently, instead of Eqs.\ (\ref{kappa}) and (\ref{tau})
we can express the diffusive thermal conductivity
in the form
\begin{eqnarray}
   \kappa_n & = & { \pi^2 T n_n
               \tau_n {\cal R}_C \over 3 m_n^\ast }, \quad
    \tau_n= { {\cal R}_C \over \nu_{nn} + \nu_{np}}, 
\nonumber \\
      \nu_{12} & = & {2^6 m_1^\ast m_2^{\ast 2} T^2
      \over 5 m_N^2 \hbar^3} \, S_{12},
\label{superlfu} \\
     S_{nn}& = & S_{n2}^{(0)} \,K_{n2} \, {\cal R}_{n2}(y_n)
\nonumber \\
     &&   + \; 3 S_{n1}^{(0)} \,K_{n1} \,
      \left[ {\cal R}_{n1}(y_n) - {\cal R}_{n2}(y_n) \right],
\nonumber \\
     S_{np}& = & S_{p2}^{(0)} \,K_{p2} \, {\cal R}_{p2}(y_n,y_p)
\nonumber \\
     && + \; { 1 \over 2} \, S_{p1}^{(0)} \,K_{p1} \,
      \left[3{\cal R}_{p1}(y_n,y_p) - {\cal R}_{p2}(y_n,y_p) \right].
\nonumber
\end{eqnarray}

The new factor ${\cal R}_C(y_n)$ is analyzed easily
in the standard manner. It is equal to 1
in the absence of superfluidity, and behaves as
${\cal R}_C(y_n) \approx 3 \sqrt{2} \, (y_n/\pi)^{3/2} \,
\exp(-y_n)$ in the limit
$y_n \to \infty$. We have calculated this factor
numerically on a dense grid of $y_n$. These results
and the asymptotes are accurately fitted as
%
%
\begin{eqnarray}
       {\cal R}_C(y_n) & =
       & \left[ 0.647 + \sqrt{(0.353)^2 + 0.109 \, y_n^2}
       \right]^{1.5} \,
\nonumber \\
      && \times \;
       \exp \left[ 1.39 - \sqrt{(1.39)^2+y_n^2} \, \right].
\label{RCsingle}
\end{eqnarray}
The maximum fit error does not exceed 1\%.

\section{Results and Discussion}
\label{s4}
Let us summarize the results of Sects.\ \ref{s2} and \ref{s3}
and present practical formulae for
the neutron thermal conductivity $\kappa_n$
in the neutron star cores:
\begin{eqnarray}
      \kappa_n & = &
      {\pi^2 k_{\rm B}^2 T n_n \tau_n {\cal R}_C (y_n) C \over
       3 m^\ast_n}
\nonumber \\
      & \approx & 7.2 \times 10^{23} \, T_8  \,
             {\cal R}_C^2(y_n) \left( \frac{m_n}{m_n^\ast} \right)
\nonumber \\
        && \times \;     \left( { 10^{15} {\rm s}^{-1} \over
             \nu_{nn} + \nu_{np} } \right)
             \left( \frac{n_n}{n_0} \right)
             \hspace{2mm}
             {\rm ergs \: cm^{-1} \: s^{-1} \: K^{-1}}.
\label{kapsupf}
\end{eqnarray}
Here,  $n_0 = 0.16 ~ {\rm fm}^{-3}$
is the normal nucleon density,
$T_8$ is temperature in units of $10^8$ K,
$k_{\rm B}$ is the Boltzmann constant
(now presented explicitly),
$C$=1.2 is the correction factor discussed in Sect.\ \ref{s27};
${\cal R}_C(y_n)$ is the
superfluid reduction factor. The latter factor depends
on the neutron gap parameter $y_n$ and is given by
Eq.\ (\ref{RCsingle}).
The gap parameter $y_n$ is determined
either by Eq.\ (\ref{delta0}) for singlet-state neutron pairing
or by Eq.\ (\ref{delta1}) for triplet-state neutron pairing
as explained in Sect.\ \ref{s3}. Let us remind that we consider
the diffusive thermal conductivity and do not analyse
convective heat transport in superfluid matter.

According to Eqs.\ (\ref{nu_through_S}) and (\ref{Reduc}),
the $nn$ and $np$ collision frequencies in Eq.\
(\ref{kapsupf})
are
\begin{eqnarray}
      \nu_{nn} & = & {2^6 m_n^{\ast 3} k_{\rm B}^2 T^2
       \over 5 m_N^2 \hbar^3} \, S_{nn}
       \approx
       3.48 \times 10^{15}
      \, \left( {m_n^\ast \over m_n} \right)^3 \, T^2_8 \,
\nonumber \\
   & & \times \;
      \left\{  S_{n2}^{(0)} \,K_{n2} \, {\cal R}_{n2}(y_n) \right.
\nonumber \\
      && + \; \left. \; 3 S_{n1}^{(0)} \,K_{n1}
      \left[ {\cal R}_{n1}(y_n) - {\cal R}_{n2}(y_n) \right] \right\}
      \; \; {\rm s}^{-1},
\nonumber \\
      \nu_{np} & = & {2^6 m_n^\ast m_p^{\ast 2} k_{\rm B}^2 T^2
      \over 5 m_N^2 \hbar^3} \, S_{np}
\nonumber \\
      & \approx &  3.48 \times 10^{15} \,
      \left( {m_n^\ast \over m_n} \right) \,
      \left( {m_p^\ast \over m_p} \right)^2 \, T^2_8 
\nonumber \\
      &  & \times \;
      \left\{ S_{p2}^{(0)} \,K_{p2} \, {\cal R}_{p2}(y_n,y_p)
      + 0.5 \, K_{p1} \right.
\nonumber \\
      && \times \; \left. 
        S_{p1}^{(0)} 
      \left[3{\cal R}_{p1}(y_n,y_p) - {\cal R}_{p2}(y_n,y_p) \right] \right\}
      \; \; {\rm s}^{-1}.
\label{tausupf}
\end{eqnarray}
Here, $S_\alpha^{(0)}$ ($\alpha$ = $n1$, $n2$, $p1$, $p2$)
is a normalized transition probability 
integrated over
momentum transfers of colliding particles
(Sects.\ \ref{s24}--\ref{s26})
for free nucleons [expressed in mb and fitted by
Eqs.\ (\ref{S_0fit})]; $K_\alpha$ describes the
in-medium effects on the squared matrix element
[the fits are given by Eqs.\ (\ref{K_fit})]; ${\cal R}_\alpha$
is a corresponding superfluid reduction factor whose fits are
given by Eqs.\ (\ref{Rn_fit})--(\ref{Rp2_fit}).
Thus we obtained a simple
description of the diffusive neutron conductivity
in the density range from
$0.5 \rho_0$ to $3 \rho_0$ (see Sect.\ \ref{s2}).
Although we have employed the Bonn model of $N\!N$ interaction,
we hope that the results could be used, at least semi-quantitatively,
for a wide class of equations of state if one takes
effective nucleon masses $m_N^\ast$, Fermi momenta
$k_{{\rm F}\!N}$, and superfluid gaps
appropriate to these equations of state.
Moreover, our results are presented in the form
which enables one to include the in-medium effects
from more elaborated (future) calculations of the
nucleon quasiparticle transition probabilities.
It would be sufficient to recalculate
four in-medium correction factors $K_\alpha$
[defined in Eq.\ (\ref{SK})] while all other quantities
in Eqs.\  (\ref{kapsupf})--(\ref{tausupf}) will remain
untouched.

\begin{figure}[t]
\begin{center}
  \leavevmode
\includegraphics[height=85mm,bb=-1 -1 346 346,clip]{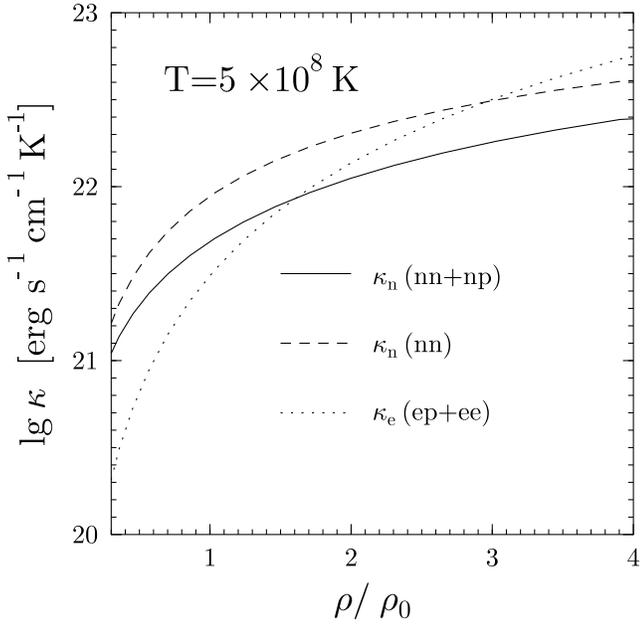}
\end{center}
\caption[]{
  Thermal conductivity of neutrons and
  electrons versus mass density
  (in units of $\rho_0 = 2.8 \times 10^{14}$ g cm$^{-3}$)
  of non-superfluid $npe$ matter in the model
  of free particles at $T=5 \times 10^8$ K;
  $m_N^\ast = m_N$.
}
\label{f1}
\end{figure}

\begin{figure}[t]
\begin{center}
\leavevmode
\includegraphics[height=85mm,bb=-1 -1 346 346,clip]{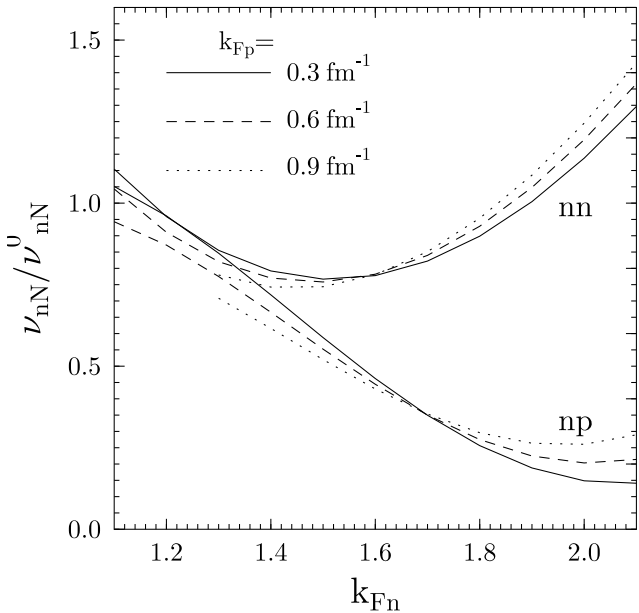}
\end{center}
\caption[]{
Ratios of in-medium to in-vacuum $nn$ and $np$ collision frequencies
versus $k_{\rm Fn}$ at $k_{\rm Fp}$=
0.3 fm$^{-1}$ (solid lines), 0.6 fm$^{-1}$ (dashes)
and 0.9 fm$^{-1}$ (dots);
$m_N^\ast = 0.8 \, m_N$.
}
\label{f2}
\end{figure}

\begin{figure}[t]
\begin{center}
\leavevmode
\includegraphics[height=85mm,bb=-1 -1 346 346,clip]{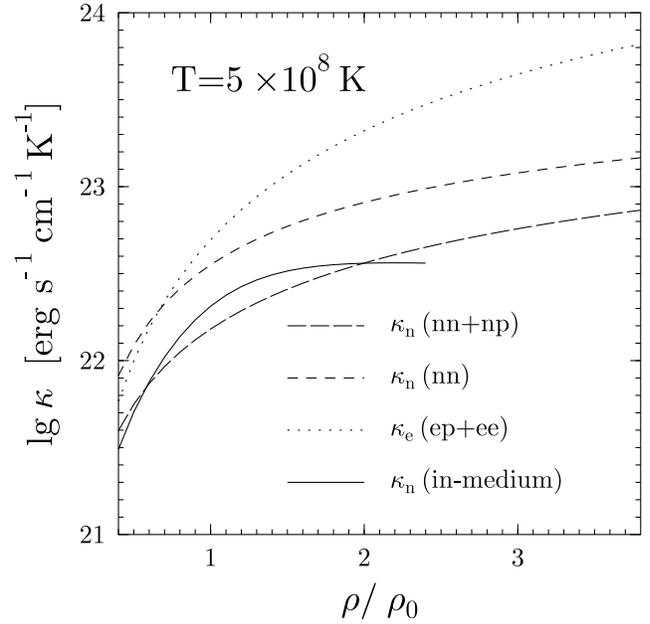}
\end{center}
\caption[]{
Thermal conductivity of neutrons and electrons
versus mass density in non-superfluid  $npe$ matter
with a moderately stiff equation of state (Prakash et al., 
\cite{Petal88}, Page \& Applegate, \cite{PA92}),
$m_N^\ast = 0.7 \, m_N$
at $T=5 \times 10^8$ K. Solid curve is calculated
with the in-medium $nn$ and $np$ scattering cross sections
while other neutron-conductivity curves
are obtained with the in-vacuum
cross sections.
}
\label{f3}
\end{figure}

\begin{figure}[t]
\begin{center}
\leavevmode
\includegraphics[height=85mm,bb=-1 -1 346 346,clip]{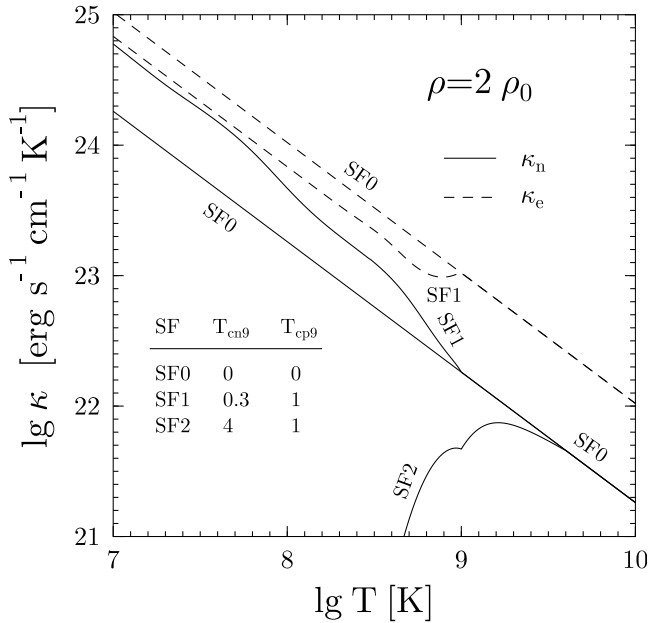}
\end{center}
\caption[]{
Diffusive thermal conductivity of neutrons (solid lines) and electrons
(dashes) versus temperature
at $\rho = 2 \rho_0$ for the same
equation of state as in Fig.\ \ref{f3}. The curves marked SF0
refer to the non-superfluid case; the curves marked SF1
imply $T_{cn}= 3 \times 10^8$ K, $T_{cp} = 10^9$ K,
while the curve SF2 corresponds to
$T_{cn}= 4 \times 10^9$ K, $T_{cp} = 10^9$ K (singlet--state 
proton and triplet--state
neutron superfluidities).
The in-medium effects in the scattering
cross sections are not included.
The dashed curve SF1 is equally valid for the case SF2.
}
\label{f4}
\end{figure}

The results are illustrated in Figs. \ref{f1}--\ref{f4}.
Figure  \ref{f1} presents the neutron thermal
conductivity versus density
at $T= 5 \times 10^8$ K for non-superfluid matter.
Here, we use the simplified model
of free degenerate neutrons, protons and electrons
(e.g., Shapiro \& Teukolsky, \cite{ST83}). This model yields an
extremely soft equation of state, with very low fraction
(about 0.6\% at $\rho=\rho_0$) of protons and electrons.
The effective masses
of nucleons are set equal to their bare masses,
and no in-medium effects are included in the $nn$ and
$np$ collision frequencies. The solid curve
depicts the full neutron conductivity, which takes
account of $nn$ and $np$ collisions. The dashed line
shows the contribution from $nn$ collisions alone.
The $nn$ and $np$
collisions are seen to be of comparable importance
despite the extreme smallness of the proton fraction.
This is related to the fact that the {\it np} collisions
occur at smaller laboratory energies and scattering angles
than the {\it nn} ones, while the $N\!N$ differential cross
sections tend to grow with decrease of both these parameters.
We present also the electron conductivity (the dotted line)
calculated using the results of Gnedin \& Yakovlev (\cite{GY95}).

Figure \ref{f2} demonstrates the significance of in-medium
effects. It displays
the ratios of in-medium to in-vacuum $nN$ collision
frequencies versus Fermi wavenumber of neutrons
$k_{\rm Fn}$ (notice that $k_{\rm Fn} \approx 1.7$ fm$^{-1}$
at $\rho = \rho_0$) at three values of the proton Fermi wavenumber,
$k_{\rm Fp}=0.3$, 0.6 and 0.9 fm$^{-1}$.
Let us emphasize again that we use one particular
model of $N\!N$ interaction in medium 
which is valid, strictly speaking,
for symmetric nuclear matter as described in Sect.\ \ref{s26}.
The in-medium effects are seen to be significant,
especially for $np$ collisions, and determined mostly by $k_{\rm Fn}$.
For instance, at $k_{\rm Fn} \approx 2$ fm$^{-1}$ the in-medium
effects reduce $\nu_{np}$ by a factor of 3--6.
The dependence of the ratios in question on
the proton fraction (on $k_{\rm Fp}$) appears rather weak.

Figure \ref{f3} shows the neutron and electron thermal conductivities
versus density at $T = 5 \times 10^8$ K for non-superfluid
matter described by a moderately stiff equation of state
proposed by Prakash et al.\  (\cite{Petal88}) [the
version with the compression modulus $K_0 = 180$ MeV,
and the same model function $F(u)$ as in Page \& Applegate (\cite{PA92})].
This equation of state is much more
realistic than
the model of free particles employed in Fig.\ \ref{f1};
it yields much larger fractions of protons and electrons
(e.g., about 6\% at $\rho = \rho_0$).
The effective nucleon masses are put equal to
$0.7 \, m_N$. The solid curve takes into account
the contribution from $nn$ and $np$ collision
frequencies evaluated with the in-medium cross sections (Fig.\ \ref{f2}).
The curve is plotted up to $\rho \approx 3 \, \rho_0$,
in the density range  where the in-medium scattering
cross sections are available (Sect.\ \ref{s26}).
Other neutron-conductivity curves are calculated
with the in-vacuum cross sections
[$K_\alpha =1$ in Eqs.\ (\ref{tausupf})].
In contrast to Fig.\ \ref{f1}, the electron thermal conductivity
exceeds the neutron one at all densities, which is explained by the much
higher fraction of electrons for the adopted equation of state.
Comparing Figs.\ \ref{f1} and \ref{f3} one can deduce that the
electron conductivity is more sensitive to
the equation of state than the neutron one. This is because
the number density of electrons varies more strongly
with the equation of state than the number density of neutrons.
In fact, the variation of the neutron conductivity
is mostly caused by the difference in effective nucleon masses
between the two models.

Finally, Fig.\ \ref{f4} demonstrates the effects
of neutron and proton superfluidity.
It shows the temperature dependence of the diffusive thermal conductivity
for the same model of matter as in Fig.\ \ref{f3}
at $\rho = 2 \rho_0$. Solid and dashed lines display the
neutron and electron thermal conductivity, respectively.
The curves marked as SF0 correspond to non-superfluid matter;
the curves SF1 correspond to the model in which
$T_{cn}=3 \times 10^8$ K, $T_{cp}   = 10^9$ K , i.e.,
the proton superfluidity is stronger than the neutron one.
Finally, the curve SF2 is for the case
$T_{cn}=4 \times 10^9$ K, $T_{cp} = 10^9$ K,
in which the neutron superfluidity is stronger.
The proton superfluidity is assumed to be of singlet type
[the dependence of the gap on the temperature is given 
by Eq.\ (\ref{delta0})] while the superfluidity of neutrons is taken to
be of triplet type with the gap made artificially 
independent of the direction of
particle momentum [the temperature dependence of the gap is given 
by Eq.\ (\ref{delta1})].
Note that the electron thermal conductivity
is independent of superfluid state of neutrons
and is the same for models SF1 and SF2.
All the solid curves are
obtained with in-vacuum cross sections.
In the absence of superfluidity, both conductivities,
$\kappa_n$ and $\kappa_e$, behave as $T^{-1}$.
When temperature falls below $T_{cn}$ or $T_{cp}$,
the conductivity depends on $T$ in a more complicated manner.

If $T_{cn} > T_{cp}$
(curve SF2, dashed curve SF1 is equally valid for
the superfluid model SF2), and the temperature falls below $T_{cn}$
the electron conductivity remains unaffected
but the diffusive neutron thermal conductivity
drops exponentially due to sharp decrease of the density of
heat carriers --- neutron quasiparticles whose momenta
are sufficiently close to the neutron Fermi momentum.
When $T$ falls below $T_{cp}$
the electron thermal conductivity
undergoes a slight reduction associated mainly
with variation in the plasma screening length
(the proton screening
which is efficient at $T>T_{cp}$ dies out leaving alone
the weaker electron screening).

If $T_{cp} > T_{cn}$ (curves SF1),
the proton superfluidity occurs first with decreasing $T$.
It switches off the $np$ collisions and thus enhances  $\kappa_n$.
When the temperature reaches
$T_{cn}$, the neutron conductivity continues to grow
over the non-superfluid one.
At further decrease of $T$
the neutron conductivity is determined by competition of two
factors: the suppression due to the
decline of the density  of the heat carriers, and the enhancement
due to lowering the $nn$ collision frequency
(while the $np$ collisions are almost switched off).
If $T \ll T_{cn}$ both factors nearly cancel each other out,
and the conductivity, although deviates somewhat
from the non-superfluid one,  but reproduces the standard temperature
dependence $T^{-1}$.
Analogous effect was demonstrated by Pethick et al.\ (\cite{PSB77})
in  the case of superfluid $^3$He at $T\ll T_c$.

We see that the nucleon superfluidity affects the neutron thermal
conductivity in different ways, producing either
enhancement or suppression, which is incorporated easily
in computations using Eqs.\ (\ref{kapsupf}) and (\ref{tausupf}).

We have considered thermal conductivity of
matter composed of neutrons, protons, electrons, and 
muons at $\rho \sim \rho_0$. Let us discuss briefly
thermal conductivity in hyperonic matter which may
appear at higher $\rho$ (e.g., Balberg et al.\ \cite{blc99}).
It may contain $\Sigma^-$, $\Lambda$ and
other hyperons. Theoretical knowledge of hyperon
interaction is rather uncertain but numerous models
of hyperonic matter give qualitatively the same results.
$\Sigma^-$ and $\Lambda$ hyperons appear at about
the same density. Their relative fractions increase
sharply with growing $\rho$ while the fractions of
electrons and muons become much lower than in the absence 
of hyperons. 
The fraction
of $\Lambda$ hyperons is typically larger than that of $\Sigma^-$
hyperons.
Qualitatively, electrons and muons
are replaced by $\Sigma^-$ hyperons while neutrons are  
are partially replaced by  $\Lambda$ hypeons. Both replacements
have important consequences for thermal conduction.

It is likely that thermal conductivities
limited by Coulomb and nuclear interactions
will remain decoupled. 
Lowering the electron and muon fractions reduces
the efficiency of the electron and muon Coulomb
thermal conductivities. These conductivities can be
determined easily using the results of Gnedin \&
Yakovlev (\cite{GY95}). On the other hand, the efficiency
of thermal conductivity, limited by strong interactions, 
becomes higher. Neutrons and $\Lambda$
hyperons may become the main heat carriers
of comparable importance. Their
thermal conductivity will be determined by collisions
of these particles with themselves and with other
strongly interacting particles (protons, $\Sigma^-$, 
etc.). Their effective collision frequencies
are likely to be coupled. The theory of thermal
conductivity of neutrons and $\Lambda$ hyperons
may be constructed using the formalism of
Flowers \& Itoh (\cite{FI79}) and Anderson et al.\ (\cite{Aetal87}).
The main problem would be to calculate the scattering
cross sections of strongly interacting particles of various species
taking into account many-body effects. 
Hyperons, like nucleons, may be in superfluid state
(e.g., Balberg et al.\ \cite{blc99}) forming multicomponent
superfluid. The effects of superfluidity on the diffusive
thermal conductivity can be included in the same manner
as in Sect.\ 3. Consideration of convective heat
transport would be even more complicated than in nucleon matter
(see above). Therefore, thermal conduction in hyperonic matter
remains an open problem. It deserves a separate study which goes
far beyond the scope of present paper. 

\section{Conclusions}
\label{s5}
We have calculated the thermal conductivity of neutrons
due to $nn$
and $np$ collisions in the neutron
star cores. The results are valid for densities $\rho$ from
$0.5 \rho_0$ to $3 \rho_0$ (see Sect.\ \ref{s2}).
We have included the effects of
possible superfluidity of neutrons and protons
(Sect.\ \ref{s3}) and obtained practical expressions
(Sect.\ \ref{s4}) for
the diffusive thermal conductivity
valid for a wide class of models of dense
matter. The results can be generalized to the case
in which hyperons are
present in dense matter along with nucleons.

Our results, combined with those of
Gnedin \& Yakovlev (\cite{GY95})
for the electron thermal conductivity, provide a
description of the transport properties in the neutron star cores.
This thermal conductivity is needed for numerical
simulations of cooling
of young neutron stars with non--isothermal core
(age up to $10-100$ years).

\acknowledgements
We are grateful to A.D.\ Kaminker and 
D.A.\ Varshalovich for useful comments,
to V.A.\ Kudryavtsev for fruitful discussions,
and also to R.\ Machleidt and G.Q.\ Li for providing us with the
$N\!N$ scattering cross sections. We acknowledge
useful remark of anonymous referee.
Two of us (DAB and DGY) are pleased to acknowledge excellent working
conditions at Copernicus Astronomical Center, Warsaw.
This work was supported in part by the
RFBR (grant No.\ 99-02-18099), INTAS (grant No.\ 96-0542),
KBN (grant No.\ 2 P 03D 01413), and NSF (grant No.\ PHY99-07949).



\begin{thebibliography}{}


\bibitem[1987]{Aetal87}
    Anderson R.H., Pethick C.J., Quader K.F., 1987,
    Phys.\ Rev.\  B35, 1620

\bibitem[1999]{blc99}
    Balberg B., Lichtenstadt I, Cook G.B., 1999,
    ApJS 121, 515

\bibitem[1971]{BBP71}
    Baym G., Bethe H.A., Pethick C.J., 1971,
    Nucl.\ Phys.\  A175, 225

\bibitem[1991]{BP91}
    Baym G., Pethick C., 1991, Landau Fermi--Liquid Theory,
    John Wiley, New York

\bibitem[1977]{BPS77}
    Bhattacharyya P., Pethick C.J., Smith H., 1977,
    Phys.\ Rev.\ B15, 3367

\bibitem[1979]{FI79}
    Flowers E., Itoh N., 1979,
    ApJ  230, 847 

\bibitem[1981]{FI81}
    Flowers E.,  Itoh N., 1981
    ApJ  250, 750 

\bibitem[1995]{GY95}
    Gnedin O.Y., Yakovlev D.G., 1995,
    Nucl.\ Phys.\ A582, 697 

\bibitem[1968]{HJeta68}
    H{\o}jgaard Jensen H., Smith H., Wilkins J.W., 1968,
    Phys.\ Lett.\  A27, 532;  
    1969, Phys. Rev.  185, 323 

\bibitem[1994]{lvpp94}
    Lattimer J.M., Van Riper K.A., Prakash M.,
    Prakash M., 1994, ApJ  425, 802 

\bibitem[1994a]{LYdurca}
    Levenfish K.P., Yakovlev D.G., 1994,
    Astron.\ Lett.\  20, 43 

\bibitem[1994b]{LYheat}
    Levenfish K.P., Yakovlev D.G., 1994,
    Astron.\ Rep.\  38, 247 

\bibitem[1993]{LM93}
   Li G.Q., Machleidt R., 1993,
   Phys.\ Rev.\  C48, 1702 

\bibitem[1994]{LM94}
   Li G.Q., Machleidt R., 1994,
   Phys.\ Rev.\  C49, 566 

\bibitem[1993]{Letal93}
    Lorenz C.P., Ravenhall D.G., Pethick C.J., 1993,
    Phys.\ Rev.\ Lett.\  70, 379 

\bibitem[1987]{Metal87}
    Machleidt R., Holinde K., Elster Ch., 1987,
    Phys.\ Rep.\  149, 1  

\bibitem[1992]{PA92}
    Page D., Applegate J.H., 1992,
    ApJ  394, L17 

\bibitem[1977]{PSB77}
    Pethick C.J., Smith H., Bhattacharyya P., 1977,
    Phys.\ Rev.\ B15, 3384 

\bibitem[1995]{PR95}
    Pethick C.J., Ravenhall D.G., 1995,
    Ann.\ Rev.\ Nucl.\ Particle Sci.\  45, 429 

\bibitem[1991]{P91}
    Pines D., 1991. In: J. Ventura and D. Pines (eds.) 
    Neutron Stars: Theory and Observation.
    Kluwer Academic Publisher, Dordrecht  p.\ 57,


\bibitem[1988]{Petal88}
    Prakash M., Ainsworth T.L., Lattimer J.M., 1988,
    Phys.\ Rev.\ Lett.\  61, 2518 

\bibitem[1994]{Setal94}
    Sedrakian A.D., Blaschke D., R$\ddot{\rm o}$pke G.,
    Schulz H., 1994,
    Phys.\ Lett.\   B338, 111 

\bibitem[1983]{ST83}
    Shapiro S.L., Teukolsky S.A., 1983,
    Black Holes, White Dwarfs, and Neutron Stars,
    Wiley-Interscience, New York 

\bibitem[1970]{SB70}
    Sykes J., Brooker G.A., 1970,
    Ann.\ Phys.\  56, 1  

\bibitem[1993]{TT93}
   Takatsuka T., Tamagaki R., 1993,
   Progr.\ Theor.\ Phys.\ Suppl.  112, 27 

\bibitem[1990]{TT90}
    Tilley D.R., Tilley J., 1990, Superfluidity and
    Superconductivity, IOP Publishing, Bristol 

\bibitem[1993]{Uetal93}
    Umeda H., Shibazaki N.,  Nomoto K., Tsuruta S., 1993,
    ApJ  408, 186 

\bibitem[1991]{VR91}
    Van Riper K.A., 1991,
    ApJS  75, 449 

\bibitem[1993]{Wetal93}
    Wambach J., Ainsworth T.L., Pines D., 1993,
    Nucl.\ Phys.\  A555, 128 (WAP)

\bibitem[1995]{YL95}
    Yakovlev D.G., Levenfish K.P., 1995,
    A\&A  297, 717 

\bibitem[1999]{yls99}
    Yakovlev D.G., Levenfish K.P., Shibanov Yu.A., 1999,
    Physics--Uspekhi  42, 737 

\bibitem[1960]{Z60}
    Ziman J.M., 1960,
    Electrons and Phonons,
    Oxford University Press, Oxford 


\end{thebibliography}
\end{document}